\documentclass[twoside]{article}
\usepackage{qic,epsfig}
\usepackage{amsmath}

 \usepackage[square,numbers,sort&compress]{natbib}

\newtheorem{mytheorem}{Theorem}
\newtheorem{fact}{Fact}
\newtheorem{proposition}{Proposition}
\newtheorem{problem}{Problem}

\newcommand{\ket}[1]{{|#1\rangle}}
\newcommand{\bra}[1]{{\langle#1|}}
\newcommand{\braket}[2]{{\langle#1|#2\rangle}}
\newcommand{\ketbra}[2]{| #1 \rangle\langle #2 |}

\newcommand{\be}{\begin{equation}}
\newcommand{\ee}{\end{equation}}
\newcommand{\bea}{\begin{eqnarray}}
\newcommand{\eea}{\end{eqnarray}}

\newcommand{\tr}{\textrm{tr}}

\newcommand{\poly}{\textrm{poly}}

\newcommand{\rank}{\textrm{rank}}

\def\opone{\leavevmode\hbox{\small1\kern-3.8pt\normalsize1}}

\newcommand{\dima}{M}
\newcommand{\dimb}{N}

\newcommand{\hermops}{\mathbf{H}_{\dima, \dimb}}

\newcommand{\densops}{\mathcal{D}_{\dima,\dimb}}
\newcommand{\densopsgen}[1]{\mathcal{D}(#1)}
\newcommand{\sep}{\mathcal{S}_{\dima, \dimb}}
\newcommand{\ent}{\mathcal{E}_{\dima, \dimb}}
\newcommand{\CMotimesCN}{\mathbf{C}^{\dima}\otimes\mathbf{C}^{\dimb}}

\newcommand{\conv}{\mathrm{conv}}

\newcommand{\n}{\dima^2\dimb^2}

\newcommand{\A}{\mathrm{A}}
\newcommand{\B}{\mathrm{B}}

\newcommand{\D}{\mathrm{D}}
\newcommand{\Y}{\mathrm{Y}}
\newcommand{\No}{\mathrm{N}}

\renewcommand{\P}{\mathrm{P}}
\newcommand{\NP}{\mathrm{NP}}

\newcommand{\NPCK}{\mathrm{NPC_\K}}
\newcommand{\NPCT}{\mathrm{NPC_\T}}

\newcommand{\K}{\mathrm{K}}
\newcommand{\T}{\mathrm{T}}

\newcommand{\col}{\mathrm{col}}

\begin{document}
\setlength{\textheight}{8.0truein}    

\runninghead{COMPUTATIONAL COMPLEXITY OF THE QUANTUM SEPARABILITY
PROBLEM}
            {LAWRENCE M. IOANNOU}

\normalsize\textlineskip
\thispagestyle{empty}
\setcounter{page}{1}

\copyrightheading{0}{0}{2003}{000--000}

\vspace*{0.88truein}

\alphfootnote

\fpage{1}

\centerline{\bf
COMPUTATIONAL COMPLEXITY OF} \vspace*{0.035truein} \centerline{\bf
THE QUANTUM SEPARABILITY PROBLEM} \vspace*{0.37truein}
\centerline{\footnotesize
LAWRENCE M. IOANNOU} \vspace*{0.015truein}
\centerline{\footnotesize\it Department of Applied Mathematics and
Theoretical Physics, University of Cambridge, Wilberforce Road}
\baselineskip=10pt \centerline{\footnotesize\it Cambridge,
Cambridgeshire, CB3 0WA, United Kingdom}
\vspace*{0.225truein} \publisher{(received
date)}{(revised date)}

\vspace*{0.21truein}

\abstracts{
Ever since entanglement was identified as a computational and
cryptographic resource, researchers have sought efficient ways to
tell whether a given density matrix represents an unentangled, or
\emph{separable}, state.  This paper gives the first systematic and
comprehensive treatment of this (bipartite) quantum separability
problem, focusing on its deterministic (as opposed to randomized)
computational complexity.  First, I review the one-sided tests for
separability, paying particular attention to the semidefinite
programming methods.  Then, I discuss various ways of formulating
the quantum separability problem, from exact to approximate
formulations, the latter of which are the paper's main focus. I then
give a thorough treatment of the problem's relationship with the
complexity classes NP, NP-complete, and co-NP.  I also discuss
extensions of Gurvits' NP-hardness result to strong NP-hardness of
certain related problems.  A major open question is whether the
NP-contained formulation (QSEP) of the quantum separability problem
is Karp-NP-complete; QSEP may be the first natural example of a
problem that is Turing-NP-complete but not Karp-NP-complete.
Finally, I survey all the proposed (deterministic) algorithms for
the quantum separability problem, including the bounded search for
symmetric extensions (via semidefinite programming), based on the
recent quantum de Finetti theorem \cite{DPS02,DPS04,qphCKMR06}; and
the entanglement-witness search (via interior-point algorithms and
global optimization) \cite{ITCE04,IT06}. These two algorithms have
the lowest complexity, with the latter being the best under advice
of asymptotically optimal point-coverings of the sphere.
 }{}{}

\vspace*{10pt}

\vspace*{3pt}
\communicate{to be filled by the Editorial}

\vspace*{1pt}\textlineskip    

\section{Introduction}\label{ChapterIntro}

If a $d$-dimensional quantum physical system can be physically
partitioned into two subsystems (denoted by superscripts $\A$ and
$\B$) of dimensions $\dima$ and $\dimb$, such that $d=\dima\dimb$,
then the pure state $\ket{\psi}$ of this total system may be
$\emph{separable}$, which means
$\ket{\psi}=\ket{\psi^{\A}}\otimes\ket{\psi^\B}$, for
$\ket{\psi^\A}\in\mathbf{C}^\dima$ and
$\ket{\psi^\B}\in\mathbf{C}^\dimb$ and where ``$\otimes$'' denotes
the Kronecker (tensor) product. Without loss of generality, assume
$\dima\leq \dimb$ (except in Section
\ref{subsubsec_StrongNPHness}).  If $\ket{\psi}$ is not separable,
then it is \emph{entangled} (with respect to that particular
partition).

Denote  by $\densopsgen{V}$ the set of all density operators
mapping complex vector space $V$ to itself; let
$\densops:=\densopsgen{\CMotimesCN}$. The \emph{maximally mixed
state} is $I_{\dima,\dimb}:=I/\dima\dimb$, where $I$ denotes the
identity operator.  
A pure state $\ket{\psi}$ is separable if and only if
$\tr_\B(\ketbra{\psi}{\psi})$ is a pure state, where ``$\tr_\B$''
denotes the partial trace with respect to subsystem $\B$; a pure
state is called \emph{maximally entangled} if
$\tr_\B(\ketbra{\psi}{\psi})$ is the maximally mixed state
$I/\dima$ in the space of density operators on the $\A$-subsystem
$\densopsgen{\mathbf{C}^\dima}$. Thus, the mixedness of
$\tr_\B(\ketbra{\psi}{\psi})$ is some ``measure'' of the
entanglement of $\ket{\psi}$.

A mixed state $\rho\in\densops$ is \emph{separable} if and only if
it may be written $\rho=\sum_{i=1}^k p_i\rho^\A_i\otimes\rho^\B_i$
with $p_i\geq 0$ and $\sum_ip_i=1$, and where
$\rho^\A_i\in\densopsgen{\mathbf{C}^\dima}$ is a (mixed or pure)
state of the $A$-subsystem (and similarly for
$\rho^\B_i\in\densopsgen{\mathbf{C}^\dimb}$); when $k=1$, $\rho$
is a \emph{product state}. Let $\sep\subset\densops$ denote the
separable states; let $\ent:=\densops\setminus\sep$ denote the
entangled states.  The following fact will be used several times
throughout this work: \vspace*{12pt}
\begin{fact}[\cite{Hor97}]\label{fact_FiniteDecompOfSepState}  If $\sigma\in\sep$, then $\sigma$ may be written as
a convex combination of $\n$ pure product states, that is,
\begin{eqnarray}
\sigma = \sum_{i=1}^{\n}p_i
\ketbra{\psi^\A_i}{\psi^\A_i}\otimes\ketbra{\psi^\B_i}{\psi^\B_i},
\end{eqnarray}
where $\sum_{i=1}^{\n}p_i=1$ and $0\leq p_i\leq 1$ for all
$i=1,2,\ldots,\n$.
\end{fact}
\vspace*{12pt} \noindent Recall that a set of points
$\{x_1,\ldots,x_j\}\subset\mathbf{R}^n$ is \emph{affinely
independent} if and only if the set
$\{x_2-x_1,x_3-x_1,\ldots,x_j-x_1\}$ is linearly independent in
$\mathbf{R}^n$.  Recall also that the \emph{dimension} of
$X\subset\mathbf{R}^n$ is defined as the size of the largest
affinely-independent subset of $X$ minus 1.  Fact
\ref{fact_FiniteDecompOfSepState} is based on the well-known
theorem of Carath{\'{e}}odory that any point in a compact convex
set $X\subset\mathbf{R}^n$ of dimension $k$ can be written as a
convex combination of $k+1$ affinely-independent extreme points of
$X$.

\vspace*{12pt}
\begin{definition}[Formal quantum separability problem]\label{def_FormalQuSep} Let $\rho\in\densops$ be a mixed state.  Given the
matrix\footnote{We do not yet define how the entries of this
matrix are encoded; at this point, we assume all entries have some
finite representation (e.g. ``$\sqrt{2}$'') and that the
computations on this matrix can be done exactly.}~ $[\rho]$ (with
respect to the standard basis of
$\mathbf{C}^\dima\otimes\mathbf{C}^\dimb$) representing $\rho$,
decide whether $\rho$ is separable.
\end{definition}
\vspace*{12pt}

\subsection{One-sided tests and
restrictions}\label{sec_OneSidedTestsAndRestrictions}

Shortly after the importance of the quantum separability problem was
recognized in the quantum information community, efforts were made
to solve it reasonably efficiently.  In this vein, many one-sided
tests have been discovered. A \emph{one-sided test (for
separability)} is a computational procedure (with input $[\rho]$)
whose output can only ever imply \emph{one} of the following (with
certainty):
\begin{itemize}\item $\rho$ is
entangled (in the case of a \emph{necessary test})\item $\rho$ is
separable (in the case of a \emph{sufficient test}).
\end{itemize}

There have been many good articles (e.g. \cite{Bru02, Ter02,
qphSSLS05}) which review the one-sided (necessary) tests. As this
work is concerned with algorithms that are both necessary and
sufficient tests for separability for all $\dima$ and $\dimb$ --
and whose computer-implementations have a hope of being useful in
low dimensions -- I only review in detail the one-sided tests
which give rise to such algorithms (see Section
\ref{sec_OneSidedTestsOnSDP}).  But here is a list of popular
conditions on $\rho$ giving rise to efficient one-sided tests for
finite-dimensional bipartite separability:

\vspace{3mm}

\noindent\textbf{Necessary conditions for $\rho$ to be separable}
\begin{itemize}

\item PPT test \cite{Per96}: $\rho^{T_\B}\geq 0$, where ``$T_\B$''
denotes partial transposition

\item Reduction criterion \cite{HH99}: $\rho^\A\otimes I-\rho \geq
0$ and $I\otimes\rho^\B-\rho \geq 0$, where $\rho_\A :=
\tr_\B(\rho)$ and ``$\tr_\B$'' denotes partial trace (and
similarly for $\rho_\B$)

\item Entropic criterion for $\alpha=2$ and in the limit
$\alpha\rightarrow 1$ \cite{HHH96a}: $S_\alpha (\rho)\geq\max
\{S_\alpha (\rho_\A),S_\alpha (\rho_\B)\}$; where, for $\alpha>1$,
$S_\alpha(\rho):=\frac{1}{1-\alpha}\textrm{ln}(\tr(\rho^\alpha))$

\item Majorization criterion \cite{NK01}: $\lambda_\rho^\downarrow
\prec \lambda_{\rho^\A}^\downarrow$ and $\lambda_\rho^\downarrow
\prec \lambda_{\rho^\B}^\downarrow$, where
$\lambda_\tau^\downarrow$ is the list of eigenvalues of $\tau$ in
nonincreasing order (padded with zeros if necessary), and $x\prec
y$ for two lists of size $s$ if and only if the sum of the first
$k$ elements of list $x$ is less than or equal to that of list $y$
for $k=1,2,...,s$; the majorization condition implies
$\max\{\rank(\rho^\A), \rank(\rho^\B)\}\leq \rank(\rho)$.

\item Computable cross-norm/reshuffling criterion
\cite{qphRud02,CW03}: $||\mathcal{U}(\rho)||_1\leq 1$, where
$||X||_1:=\tr(\sqrt{X^\dagger X})$ is the trace norm; and
$\mathcal{U}(\rho)$, an $\dima^2\times\dimb^2$ matrix, is defined
on product states as $\mathcal{U}(A\otimes B):=v(A)v(B)^T$, where,
relative to a fixed basis,
$[v(A)]=(\col_1([A])^T,\ldots,\col_\dima([A])^T)^T$ (and similarly
for $v(B)$), where $\col_i([A])$ is the $i$th column of matrix
$[A]$; more generally \cite{qphHHH02}, any linear map
$\mathcal{U}$ that does not increase the trace norm of product
states may be used.

\end{itemize}

\noindent\textbf{Sufficient conditions for $\rho$ to be separable}
\begin{itemize}

\item Distance from maximally mixed state (see also
\cite{BCJLPS99}):

\begin{itemize}\item \cite{GB02}: e.g. $\tr(\rho - I_{\dima,
\dimb})^2\leq1/\dima\dimb(\dima\dimb-1)$\end{itemize}

\begin{itemize}\item \cite{ZHSL98,VT99} $\lambda_{\min}(\rho)\geq
(2+\dima\dimb)^{-1}$, where $\lambda_{\min}(\rho)$ denotes the
smallest eigenvalue of $\rho$\end{itemize}

\item When $\dima=2$ \cite{KCKL00}: $\rho=\rho^{T_A}$.
\end{itemize}

When $\rho$ is of a particular form, the PPT test is necessary and
sufficient for separability. This happens when
\begin{itemize}
\item $\dima\dimb \leq 6$ \cite{HHH96}; or

\item $\rank(\rho)\leq \dimb$ \cite{KCKL00,HLVC00}, see also
\cite{AFG01}.
\end{itemize}

The criteria not based on eigenvalues are obviously efficiently
computable; i.e. computing the natural logarithm can be done with
a truncated Taylor series, and the rank can be computed by
Gaussian elimination. That the tests based on the remaining
criteria are efficiently computable follows from the efficiency of
algorithms for calculating the spectrum of a Hermitian operator.
The method of choice for computing the entire spectra is the QR
algorithm (see any of \cite{WR71, GL96, SB02}), which has been
shown to have good convergence properties \cite{Wil68}.

In a series of articles (\cite{LS98}, \cite{KCKL00},
\cite{HLVC00}), various conditions for separability were obtained
which involve product vectors in the ranges of $\rho$ and
$\rho^{T_A}$. Any constructive separability checks given therein
involve computing these product vectors, but no general bounds
were obtained by the authors on the complexity of such
computations.

\subsection{One-sided tests based on semidefinite
programming}\label{sec_OneSidedTestsOnSDP}

Let $\hermops$ denote the set of all Hermitian operators mapping
$\mathbf{C}^\dima\otimes\mathbf{C}^\dimb$ to
$\mathbf{C}^\dima\otimes\mathbf{C}^\dimb$; thus,
$\densops\subset\hermops$. This vector space is endowed with the
Hilbert-Schmidt inner product $\langle X, Y \rangle\equiv
\tr(AB)$, which induces the corresponding norm
$||X||\equiv\sqrt{\tr(X^2)}$ and distance measure $||X-Y||$. By
fixing an orthogonal Hermitian basis for $\hermops$, the elements
of $\hermops$ are in one-to-one correspondence with the elements
of the real Euclidean space $\mathbf{R}^{\n}$.  If the Hermitian
basis is orthonormal, then the Hilbert-Schmidt inner product in
$\hermops$ corresponds exactly to the Euclidean dot product in
$\mathbf{R}^{\n}$.

Let us be more precise.  Let
$\mathcal{B}=\{X_i:i=0,1,\ldots,\n-1\}$ be an orthonormal,
Hermitian basis for $\mathbf{H}_{M,N}$, where
$X_0\equiv\frac{1}{\sqrt{MN}}I$. For concreteness, we can assume
that the elements of $\mathcal{B}$ are tensor-products of the
(suitably normalized) canonical generators of SU(M) and SU(N),
given e.g. in \cite{TNWM02}. Note $\tr(X_i)=0$ for all $i>0$.
Define $v:\hermops\rightarrow \mathbf{R}^{\dima^2\dimb^2-1}$ as
\begin{eqnarray}\label{eqn_MappingFromHermopsToRealVecs}
v(A):=\begin{bmatrix} \tr(X_1 A) \\ \tr(X_2 A) \\
\vdots \\ \tr(X_{\dima^2\dimb^2-1} A)\end{bmatrix}.
\end{eqnarray}
Via the mapping $v$, the set of separable states $\sep$ can be
viewed as a full-dimensional convex subset of
$\mathbf{R}^{\dima^2\dimb^2-1}$
\begin{eqnarray}
\{v(\sigma)\in \mathbf{R}^{\dima^2\dimb^2-1}: \sigma\in\sep\},
\end{eqnarray}
which properly contains the origin
$v(I_{\dima,\dimb})=\overline{0}\in\mathbf{R}^{\dima^2\dimb^2-1}$
(recall that there is a ball of separable states of nonzero radius
centred at the maximally mixed state $I_{\dima,\dimb}$).

Thus $\densops$ and $\sep$ may be viewed as subsets of the Euclidean
space $\mathbf{R}^{\n}$; actually, because all density operators
have unit trace, $\densops$ and $\sep$ are full-dimensional subsets
of $\mathbf{R}^{\n-1}$.  This observation aids in solving the
quantum separability problem, allowing us to apply easily
well-studied mathematical-programming tools. The following is from
the popular review article of semidefinite programming in
\cite{VB96}. \vspace*{12pt}
\begin{definition}[Semidefinite program (SDP)] Given the
vector $c\in\mathbf{R}^m$ and Hermitian matrices
$F_i\in\mathbf{C}^{n\times n}$, $i=0,1,\ldots,m$,
\begin{eqnarray}
&\textrm{minimize} \hspace{2mm}& c^Tx\\
&\textrm{subject to:}\hspace{2mm}& F(x)\geq 0,
\end{eqnarray}
where $F(x):= F_0 + \sum_{i=1}^{m}x_iF_i$.
\end{definition}
\vspace*{12pt} \noindent Call $x$ \emph{feasible} when $F(x)\geq
0$. When $c=0$, the SDP reduces to the \emph{semidefinite
feasibility problem}, which is to find an $x$ such that $F(x)\geq
0$ or assert that no such $x$ exists. Semidefinite programs can be
solved efficiently, in time $O(m^2n^{2.5})$.  Most algorithms are
iterative.  Each iteration can be performed in time $O(m^2n^{2})$.
The number of required iterations has an analytical bound of
$O(\sqrt{n})$, but in practice is more like $O(\log(n))$ or
constant.

\subsubsection{A test based on symmetric
extensions}\label{sec_DohertyEtalApproach}

Consider a separable state
$\sigma=\sum_i p_i
\ketbra{\psi^\A_i}{\psi^\A_i}\otimes\ketbra{\psi^\B_i}{\psi^\B_i}$,
and consider the following \emph{symmetric extension of $\sigma$
to $k$ copies of subsystem $\A$} ($k\geq 2$):
\begin{eqnarray}
\tilde{\sigma}_k=\sum_i p_i
(\ketbra{\psi^\A_i}{\psi^\A_i})^{\otimes
k}\otimes\ketbra{\psi^\B_i}{\psi^\B_i}.
\end{eqnarray}
\noindent The state $\tilde{\sigma}_k$ is so called because it
satisfies two properties: (i) it is symmetric (unchanged) under
permutations (swaps) of any two copies of subsystem $\A$; and (ii)
it is an extension of $\sigma$ in that tracing out any of its
$(k-1)$ copies of subsystem $\A$ gives back $\sigma$.  For an
arbitrary density operator $\rho\in\densopsgen{\CMotimesCN}$,
define a \emph{symmetric extension of $\rho$ to $k$ copies of
subsystem $\A$} as any density operator
$\rho'\in\densopsgen{(\mathbf{C}^\dima)^{\otimes
k}\otimes\mathbf{C}^\dimb}$ that satisfies (i) and (ii) with
$\rho$ in place of $\sigma$. If $\rho$ does not have a symmetric
extension to $k_0$ copies of subsystem $\A$ for some $k_0$, then
$\rho\notin\sep$ (else we could construct $\tilde{\rho}_{k_0}$).
Thus a method for searching for symmetric extensions of $\rho$ to
$k$ copies of subsystem $\A$ gives a sufficient test for
separability.

Doherty et al.\ \cite{DPS02, DPS04} showed that the search for a
symmetric extension to $k$ copies of $\rho$ (for any fixed $k$)
can be phrased as a SDP\@. This result, combined with the
``quantum de Finetti theorem'' \cite{FLV88, CFS02} that
$\rho\in\sep$ if and only if, for all $k$, $\rho$ has a symmetric
extension to $k$ copies of subsystem $\A$, gives an infinite
hierarchy (indexed by $k=2,3,\ldots$) of SDPs with the property
that, for each entangled state $\rho$, there exists a SDP in the
hierarchy whose solution will imply that $\rho$ is entangled.

Actually, Doherty et al.\ develop a stronger test, inspired by
Peres' PPT test.  The state $\tilde{\sigma}_k$, which is positive
semidefinite, satisfies a third property: (iii) it remains
positive semidefinite under all possible partial transpositions.
Thus $\tilde{\sigma}_k$ is more precisely called a \emph{PPT
symmetric extension}.  The SDP can be easily modified to perform a
search for PPT symmetric extensions without any significant
increase in computational complexity (one just needs to add
constraints that force the partial transpositions to be positive
semidefinite). This strengthens the separability test, because a
given (entangled) state $\rho$ may have a symmetric extension to
$k_0$ copies of subsystem $\A$ but may not have a PPT symmetric
extension to $k_0$ copies of subsystem $\A$ (Doherty et al.\ also
show that the $(k+1)$st test in this stronger hierarchy subsumes
the $k$th test).

The final SDP has the following form:
\begin{equation}\label{prob_DohertyEtalSDP}
\begin{array}{rlrcl}
&\textrm{minimize} \hspace{2mm} & 0 && \\
&\textrm{subject to:}\hspace{2mm}& \tilde{X}_k &\geq& 0  \\
&\hspace{2mm}&  (\tilde{X}_k)^{T_j}&\geq&0,\hspace{2mm} j\in J,
\end{array}
\end{equation}
where $\tilde{X}_k$ is a parametrization of a symmetric extension
of $\rho$ to $k$ copies of subsystem $\A$, and $J$ is the set of
all subsets of the $(k+1)$ subsystems that give rise to
inequivalent partial transposes $(\tilde{X}_k)^{T_j}$ of
$\tilde{X}_k$.  By noting that we can restrict our search to
so-called \emph{Bose-symmetric extensions}, where $(I\otimes
P)\rho'=\rho'$ for all $k!$ permutations $P$ of the $k$ copies of
subsystem $\A$ (as opposed to just extensions where $(I\otimes
P)\rho'(I\otimes P^\dagger)=\rho'$ for all permutations $P$), the
number of variables of the SDP is
$m=((d_{S_k})^2-\dima^2)\dimb^2$, where $d_{S_k}$=
${\dima+k-1} \choose k$ is the dimension of the symmetric subspace
of $(\mathbf{C}^\dima)^{\otimes k}$.  The size of the matrix
$\tilde{X}_k$ for the first constraint is $(d_{S_k})^2\dimb^2$.
The number of inequivalent partial transpositions is
$|J|=k$.\footnote{Choices are: transpose subsystem $\B$, transpose
1 copy of subsystem $\A$, transpose 2 copies of subsystem $\A$,
..., transpose $k-1$ copies of subsystem $\A$. Transposing all $k$
copies of subsystem $\A$ is equivalent to transposing subsystem
$\B$.  Transposing with respect to both subsystem $\B$ and $l$
copies of subsystem $\A$ is equivalent to transposing with respect
to $k-l$ copies of subsystem $\A$.}~~ The constraint corresponding
to the transposition of $l$ copies of $\A$, $l=1,2,...,k-1$, has a
matrix of size $(d_{S_l})^2(d_{S_{(k-l)}})^2\dimb^2$ \cite{DPS04}.
I will estimate the total complexity of this approach to the
quantum separability problem in Section
\ref{sec_DohertyEtalKonigRenner}.


\subsubsection{A test based on semidefinite
relaxations}\label{subsec_EisertsEtalApproach}

Doherty et al.\ formulate a \emph{hierarchy of necessary criteria}
for separability in terms of semidefinite programming -- each
separability criterion in the hierarchy may be checked by a SDP\@.
As it stands, their approach is manifestly a one-sided test for
separability, in that at no point in the hierarchy can one
conclude that the given $[\rho]$ corresponds to a separable state
(happily, recent results show that, for sufficiently large $k$,
the symmetric-extension test is a complete approximate
separability test; see Section \ref{sec_DohertyEtalKonigRenner}).

Eisert et al. \cite{EHGC04} formulate a \emph{necessary and
sufficient criterion} for separability as a hierarchy of SDPs.
Define the function
\begin{eqnarray}\label{eqn_DefnOfQuasiRelEntropy}
E_{d^2_2}(\rho) := \min_{x\in\sep} \tr((\rho - x)^2)
\end{eqnarray}
for $\rho\in\densops$.  As $\tr((\rho - x)^2)$ is the square of
the Euclidean distance from $\rho$ to $x$, $\rho$ is separable if
and only if $E_{d^2_2}(\rho)=0$.  The problem of computing
$E_{d^2_2}(\rho)$ (to check whether it is zero) is already
formulated as a constrained optimization. The following
observation helps to rewrite these constraints as low-degree
polynomials in the variables of the problem: \vspace*{12pt}
\begin{fact}[\cite{EHGC04}]\label{fact_EisertPurityFact}
Let $O$ be a Hermitian operator and let $\alpha\in\mathbf{R}$
satisfy $0<\alpha\leq 1$.  If $\tr(O^2)=\alpha^2$ and
$\tr(O^3)=\alpha^3$, then $\tr(O)=\alpha$ and $\rank(O)=1$ (i.e.
$O$ corresponds to an unnormalized pure state).
\end{fact}\vspace*{12pt}
\noindent Combining Fact \ref{fact_EisertPurityFact} with Fact
\ref{fact_FiniteDecompOfSepState}, the problem is equivalent to
\begin{equation}\label{prob_EisertsQSEP}
\begin{array}{rlrcl}
&\textrm{minimize} \hspace{2mm} & \tr((\rho - \sum_{i=1}^{\n}X_{i})^2) && \\
&\textrm{subject to:     }\hspace{10mm}& \tr(\sum_{i=1}^{\n}X_{i}) &=& 1  \\
&\hspace{2mm}& \tr((\tr_{j}(X_{i}))^2)&=& (\tr(X_i))^2,\hspace{2mm}\\
&&&&\textrm{for $i=1,2,\ldots,\n$ and $j\in\{\A,\B\}$}  \\
&\hspace{2mm}& \tr((\tr_{j}(X_{i}))^3)&=& (\tr(X_i))^3,\hspace{2mm}\\
&&&&\textrm{for $i=1,2,\ldots,\n$ and $j\in\{\A,\B\}$},
\end{array}
\end{equation}
where the new variables are Hermitian matrices $X_{i}$ for
$i=1,2,\ldots,\n$.  The constraints do \emph{not} require $X_{i}$
to be tensor products of \emph{unit-trace} pure density operators,
because the positive coefficients (probabilities summing to 1)
that would normally appear in the expression
$\sum_{i=1}^{\n}X_{i}$ are absorbed into the $X_{i}$, in order to
have fewer variables (i.e. the $X_i$ are constrained to be density
operators corresponding to unnormalized pure product states). Once
an appropriate Hermitian basis is chosen for $\hermops$, the
matrices $X_{i}$ can be parametrized by the real coefficients with
respect to the basis; these coefficients form the real variables
of the feasibility problem. The constraints in
(\ref{prob_EisertsQSEP}) are polynomials in these variables of
degree less than or equal to 3.\footnote{Alternatively, we could
parametrize the pure states (composing $X_i$) in
$\mathbf{C}^\dima$ and $\mathbf{C}^\dimb$ by the real and
imaginary parts of rectangularly-represented complex coefficients
with respect to the standard bases of $\mathbf{C}^\dima$ and
$\mathbf{C}^\dimb$:
\begin{equation}\label{prob_EisertsQSEPStandardBasis}
\begin{array}{rlrcl}
&\textrm{minimize} \hspace{2mm} & 0 && \\
&\textrm{subject to:     }\hspace{10mm}& \tr((\rho - \sum_{i=1}^{\n}\ketbra{\psi^{\A}_i}{\psi^{\A}_i}\otimes\ketbra{\psi^{\B}_i}{\psi^{\B}_i})^2)&=& 0  \\
&\hspace{2mm}& \tr\left(\sum_{i=1}^{\n}\ketbra{\psi^{\A}_i}{\psi^{\A}_i}\otimes\ketbra{\psi^{\B}_i}{\psi^{\B}_i}\right)  &=& 1.  \\
\end{array}
\end{equation}
This parametrization hard-wires the constraint that the
$\ketbra{\psi^{\A}_i}{\psi^{\A}_i}\otimes\ketbra{\psi^{\B}_i}{\psi^{\B}_i}$
are (unnormalized) pure product states, but increases the degree
of the polynomials in the constraint to 4 (for the unit trace
constraint) and 8 (for the distance constraint).}

Polynomially-constrained optimization problems can be approximated
by, or \emph{relaxed} to, semidefinite programs, via a number of
different approaches (see references in
\cite{EHGC04}).\footnote{For our purposes, the idea of a
relaxation can be briefly described as follows. The given problem
is to solve $\min_{x\in\mathbf{R}^n}\{p(x):\hspace{1mm}g_k(x)\geq
0, k=1,\ldots,m\}$, where
$p(x),g_i(x):\hspace{1mm}\mathbf{R}^n\rightarrow \mathbf{R}$ are
real-valued polynomials in $\mathbf{R}[x_1,\ldots,x_n]$.  By
introducing new variables corresponding to products of the given
variables (the number of these new variables depends on the
maximum degree of the polynomials $p,g_i$), we can make the
objective function linear in the new variables; for example, when
$n=2$ and the maximum degree is 3, if
$p(x)=3x_1+2x_1x_2+4x_1x_2^2$ then the objective function is
$c^Ty$ with $c=(0,3,0,0,2,0,0,0,4,0)\in\mathbf{R}^{10}$ and
$y\in\mathbf{R}^{10}$, where $10$ is the total number of monomials
in $\mathbf{R}[x_1,x_2]$ of degree less than or equal to 3. Each
polynomial defining the feasible set
$G:=\{x\in\mathbf{R}^n:\hspace{1mm}g_k(x)\geq 0, k=1,\ldots,m\}$
can be viewed similarly.  A relaxation of the original problem is
a SDP with objective function $c^Ty$ and with a (convex) feasible
region (in a higher-dimensional space) whose projection onto the
original space $\mathbf{R}^n$ approximates $G$. Better
approximations to $G$ can be obtained by going to higher
dimensions.}~ Some approaches even give an asymptotically complete
hierarchy of SDPs, indexed on, say, $i=1,2,\ldots$.  The SDP at
level $i+1$ in the hierarchy gives a better approximation to the
original problem than the SDP at level $i$; but, as expected, the
size of the SDPs grows with $i$ so that better approximations are
more costly to compute.  The hierarchy is asymptotically complete
because, under certain conditions, the optimal values of the
relaxations converge to the optimal value of the original problem
as $i\rightarrow\infty$. Of these approaches, the method of
Lasserre \cite{Las01} is appealing because a computational package
\cite{HL03} written in MATLAB is freely available. Moreover, this
package has built into it a method for recognizing when the
optimal solution to the original problem has been found (see
\cite{HL03} and references therein). Because of this feature, the
one-sided test becomes, in practice, a full algorithm for the
quantum separability problem.  However, no analytical worst-case
upper bounds on the running time of the algorithm for arbitrary
$\rho\in\densops$ are presently available.

\subsubsection{Entanglement
Measures}\label{subsec_EntanglementMeasures}

The function $E_{d^2_2}(\rho)$ defined in
(\ref{eqn_DefnOfQuasiRelEntropy}), but first defined in
\cite{VPRK97}, is also known as an \emph{entanglement measure},
which, at the very least, is a nonnegative real function defined
on $\densops$ (for a comprehensive review of entanglement
measures, see \cite{Chr05}).  If an entanglement measure $E(\rho)$
satisfies
\begin{eqnarray}\label{crit_VanSepAndPosEnt}
E(\rho)=0\hspace{2mm}\Leftrightarrow\hspace{2mm}\rho\in\sep,
\end{eqnarray}
then, in principle, any algorithm for computing $E(\rho)$ gives an
algorithm for the quantum separability problem. Note that most
entanglement measures $E$ do not satisfy
(\ref{crit_VanSepAndPosEnt}); most just satisfy
$E(\rho)=0\Leftarrow\rho\in\sep$.

A class of entanglement measures that do satisfy
(\ref{crit_VanSepAndPosEnt}) are the so-called ``distance
measures'' $E_d(\rho):= \min_{\sigma\in\sep} d(\rho,\sigma)$, for
any reasonable measure of ``distance'' $d(x,y)$ satisfying
$d(x,y)\geq 0$ and $(d(x,y)=0)\Leftrightarrow (x=y)$. If $d$ is
the square of the Euclidean distance, we get $E_{d^2_2}(\rho)$.
Another ``distance measure'' is the von Neumann relative entropy
$S(x,y):= \tr(x(\log x - \log y))$.

In Eisert et al.'s approach, we could replace $E_{d^2_2}$ by $E_d$
for any ``distance function'' $d(\rho,\sigma)$ that is expressible
as a polynomial in the variables of $\sigma$.  What dominates the
running time of Eisert et al.'s approach is the implicit
minimization over $\sep$, so using a different ``distance
measure'' (i.e. only changing the first constraint in
(\ref{prob_EisertsQSEP})) like $(\tr(\rho-\sigma))^2$ would not
improve the analytic runtime (because the degree of the polynomial
in the constraint is still 2), but may help in practice.

Another entanglement measure $E$ that satisfies
(\ref{crit_VanSepAndPosEnt}) 
 is the \emph{entanglement of formation} \cite{BDSW96}
\begin{eqnarray}
E_F(\rho):= \min_{\{p_i,\ketbra{\psi_i}{\psi_i}\}_i:\hspace{2mm}
\rho=\sum_ip_i\ketbra{\psi_i}{\psi_i}}\sum_ip_iS(\tr_{\B}(\ketbra{\psi_i}{\psi_i})),
\end{eqnarray}
where $S(\rho):=-\tr(\rho\log(\rho))$ is the von Neumann entropy.
This gives another strategy for a separability algorithm: search
through all decompositions of the given $\rho$ to find one that is
separable.  We can implement this strategy using the same
relaxation technique of Eisert et al., but first we have to
formulate the strategy as a polynomially-constrained optimization
problem. The role of the function $S$ is to measure the
entanglement of $\ketbra{\psi_i}{\psi_i}$ by measuring the
mixedness of the reduced state
$\tr_{\B}(\ketbra{\psi_i}{\psi_i})$.  For our purposes, we can
replace $S$ with any other function $T$ that measures mixedness
such that, for all $\rho\in\densops$, $T(\rho)\geq 0$ and
$T(\rho)=0$ if and only if $\rho$ is pure. Recalling that, for any
$\rho\in\densops$, $\tr(\rho^2)\leq 1$ with equality if and only
if $\rho$ is pure, the function $T(\rho):= 1 - \tr(\rho^2)$
suffices; this function $T$ may be written as a (finite-degree)
polynomial in the real variables of $\rho$, whereas $S$ could not.
Defining
\begin{eqnarray}\label{eqn_DefnMyPurityEntanglementMeasure}
E'_F(\rho):= \min_{\{p_i,\ketbra{\psi_i}{\psi_i}\}_i:\hspace{2mm}
\rho=\sum_ip_i\ketbra{\psi_i}{\psi_i}}\sum_ip_iT(\tr_{\B}(\ketbra{\psi_i}{\psi_i})),
\end{eqnarray}
we have that $E'_F$ satisfies (\ref{crit_VanSepAndPosEnt}). Using
an argument similar to the proof of Lemma 1 in \cite{Uhl98}, we
can show that the minimum in
(\ref{eqn_DefnMyPurityEntanglementMeasure}) is attained by a
\emph{finite} decomposition of $\rho$ into $\n+1$ pure states.
Thus, the following polynomially-constrained optimization problem
can be approximated by semidefinite relaxations:
\begin{equation}\label{prob_MyEisert-likeQSEP}
\begin{array}{rlrcl}
&\textrm{minimize} \hspace{2mm} & \sum_{i=1}^{\n+1}\tr(X_{i})T(\tr_{\B}(X_{i})) && \\
&\textrm{subject to:     }& \tr(\sum_{i=1}^{\n+1}X_{i}-[\rho])^2 &=& 0 \\
&\hspace{2mm}& \tr(\sum_{i=1}^{\n+1}X_{i}) &=& 1  \\
&\hspace{2mm}& \tr(X_{i}^2)&=& (\tr(X_i))^2,\hspace{2mm}\\
&&&&\textrm{for $i=1,2,\ldots,\n+1$}\\
&\hspace{2mm}& \tr(X_{i}^3)&=& (\tr(X_i))^3,\hspace{2mm}\\
&&&&\textrm{for $i=1,2,\ldots,\n+1$}.
\end{array}
\end{equation}
The above has about half as many constraints as
(\ref{prob_EisertsQSEP}), so it would be interesting to compare
the performance of the two approaches.

\subsubsection{Other tests}\label{subsubsec_OtherTests}

There are several one-sided tests which do not lead to full
algorithms for the quantum separability problem for $\sep$.
\vspace*{12pt}
\begin{definition}[Robust semidefinite program] Given the
vector $c\in\mathbf{R}^m$, Hermitian matrices
$F_i\in\mathbf{C}^{n\times n}$, $i=0,1,\ldots,m$, and vector space
$\mathcal{D}$,
\begin{eqnarray}
&\textrm{minimize} \hspace{2mm}& c^Tx\\
&\textrm{subject to:}\hspace{2mm}& F(x,\Delta)\geq 0 \textrm{, for
all $\Delta \in \mathcal{D}$,}
\end{eqnarray}
where $F(x,\Delta):= F_0(\Delta) + \sum_{i=1}^{m}x_iF_i(\Delta)$.
\end{definition}
\vspace*{12pt} Brand{\~{a}}o and Vianna \cite{BV04} have a set of
one-sided necessary tests based on deterministic relaxations of a
robust semidefinite program, but this set is not an asymptotically
complete hierarchy. The same authors also have a related
\emph{randomized} quantum separability algorithm which uses
probabilistic relaxations of the same robust semidefinite program
\cite{BV04a} (but randomized algorithms are outside of our scope).
I give their robust semidefinite program at the end of Section
\ref{subsubsec_LargeSDPPerezGarcia}, where we will see a similar
(nonrobust) SDP -- essentially, a discretization of the robust
semidefinite program -- that solves the (approximate) quantum
separability problem.

Woerdeman \cite{Woe03} has a set of one-sided tests for the case
where $\dima=2$.  His approach might be described as the
mirror-image of Doherty et al.'s: Instead of using an infinite
hierarchy of necessary criteria for separability, he uses an
infinite hierarchy of sufficient criteria.  Each criterion in the
hierarchy can be checked with a SDP.

\section{Separability as a Computable Decision
Problem}\label{ChapterSepAsDecisionProblem}

Definition \ref{def_FormalQuSep} gave us a concrete definition of
the quantum separability problem that we could use to explore some
important results.  Now we step back from that definition and, in
Section \ref{subsec_FormulatingQuSep}, consider more carefully how
we might formulate the quantum separability problem for the
purposes of computing it.  After considering exact formulations in
Section \ref{subsubsec_ExactFormulations}, we settle on
approximate formulations of the problem in Section
\ref{subsubsec_ApproxFormulations}, and give a few examples that
are, in a sense, equivalent.

In Section \ref{subsec_CompComplexity}, I discuss various aspects
of the computational complexity of the quantum separability
problem.  Section \ref{subsubsec_ReviewNPCness} contains a review
of NP-completeness theory.  In Sections \ref{subsubsec_DefnInNP}
and \ref{subsubsec_NPHardnessOfQSEP}, I give a formulation of the
quantum separability problem that is NP-complete with respect to
Turing reductions.    In Section
\ref{subsubsec_NonmembershipInCoNP}, I consider the quantum
separability problem's membership in co-NP\@.  In Section
\ref{subsubsec_StrongNPHness}, I explore the problem of strong
NP-hardness of the (approximate) quantum separability problem.
 Finally, in Section \ref{subsubsec_TowardsKarp}, I discuss the
open problem of whether the quantum separability problem is
NP-complete with respect to Karp reductions.

\subsection{Formulating the quantum separability problem}\label{subsec_FormulatingQuSep}

The nature of the quantum separability problem and the possibility
for quantum computers allows a number of approaches, depending on
whether the input to the problem is classical (a matrix
representing $\rho$) or quantum ($T$ copies of a physical system
prepared in state $\rho$) and whether the processing of the input
will be done on a classical computer or on a quantum computer. The
use of entanglement witnesses\footnote{An \emph{entanglement
witness (for $\rho$)} is defined to be any operator $A\in\hermops$
such that $\tr(A\sigma)<\tr(A\rho)$ for all $\sigma\in\sep$ and
some $\rho\in\ent$; we say that ``$A$ detects $\rho$''.  Every
$\rho\in\ent$ has an entanglement witness that detects it
\cite{HHH96}.}\label{footnote_DefnEW}~~ in the laboratory is a
case of a quantum input and very limited quantum processing in the
form of measurement of each copy of $\rho$.  The case of
more-sophisticated quantum processing on either a quantum or
classical input is not well studied (see \cite{HE02} for an
instance of more-sophisticated quantum processing on a quantum
input). For the remainder of the paper, I focus on the case where
input and processing are classical (though the algorithm in
Section \ref{sec_NewSepAlg} can be applied in an experimental
setting \cite{ITCE04,IT06}).

\subsubsection{Exact formulations}\label{subsubsec_ExactFormulations}

Let us examine Definition \ref{def_FormalQuSep}
from a computational viewpoint.  The
matrix $[\rho]$ is allowed to have real entries.  Certainly there
are real numbers that are uncomputable (e.g. a number whose $n$th
binary digit is 1 if and only if the $n$th Turing machine halts on
input $n$); we disallow such inputs.  However, the real numbers
$e$, $\pi$, and $\sqrt{2}$ are computable to any degree of
approximation, so in principle they should be allowed to appear in
$[\rho]$.  In general, we should allow any real number that can be
approximated arbitrarily well by a computer subroutine. If
$[\rho]$ consists of such real numbers (subroutines), say that
``$\rho$ is given as an approximation algorithm for $[\rho]$.'' In
this case, we have a procedure to which we can give an accuracy
parameter $\delta>0$ and out of which will be returned a matrix
$[\rho]_\delta$ that is (in some norm) at most $\delta$ away from
$[\rho]$.  Because $\sep$ is closed, the sequence
$([\rho]_{1/n})_{n=1,2,\ldots}$ may converge to a point on the
boundary of $\sep$ (when $\rho$ is on the boundary of $\sep$). For
such $\rho$, the formal quantum separability problem may be
``undecidable'' because the $\delta$-radius ball centred at
$[\rho]_\delta$ may contain both separable and entangled states
for all $\delta>0$ \cite{Myr97} (more generally, see ``Type II
computability'' in \cite{Wei87}).

If we really want to determine the complexity of deciding
membership in $\sep$, it makes sense not to confuse this with the
complexity of specifying the input.  To give the computer a
fighting chance, it makes more sense to restrict to inputs that
have finite exact representations that can be readily subjected to
elementary arithmetic operations begetting exact answers. For this
reason, we might restrict the formal quantum separability problem
to instances where $[\rho]$ consists of rational entries:
\vspace*{12pt}
\begin{definition}[Rational quantum separability problem (EXACT QSEP)]\label{def_RationalQuSep}
Let \\$\rho\in\densops$ be a mixed state such that the matrix
$[\rho]$ (with respect to the standard basis of
$\mathbf{C}^\dima\otimes\mathbf{C}^\dimb$) representing $\rho$
consists of rational entries.  Given $[\rho]$, is $\rho$
separable?
\end{definition}
\vspace*{12pt}

As pointed out in \cite{DPS04}, Tarski's
algorithm\footnote{Tarski's result is often called the
``Tarski-Seidenberg'' theorem, after Seidenberg, who found a
slightly better algorithm \cite{Sei54} (and elaborated on its
generality) in 1954, shortly after Tarski managed to publish his;
but Tarski discovered his own result in 1930 (the war prevented
him from publishing before 1948).}~ \cite{Tar51} can be used to
solve EXACT QSEP\@.  The Tarski-approach is as follows. Note that
the following first-order logical formula\footnote{Recall the
logical connectives: $\vee$ (``OR''), $\wedge$ (``AND''), $\neg$
(``NOT''); the symbol $\rightarrow$ (``IMPLIES''), in
``$x\rightarrow y$'', is a shorthand, as ``$x\rightarrow y$'' is
equivalent to ``$(\neg x) \vee y $''; as well, we can consider
``$x\vee y$'' shorthand for ``$\neg((\neg x)\wedge (\neg y))$''.
Also recall the existential and universal quantifiers $\exists$
(``THERE EXISTS'') and $\forall$ (``FOR ALL''); note that the
universal quantifier $\forall$ is redundant as ``$\forall x
\phi(x)$'' is equivalent to ``$\neg\exists x\neg\phi(x)$''.}~ is
true if and only if $\rho$ is separable:
\begin{eqnarray}\label{eqn_FOLogicalStatementOfSep}
\forall A[(\forall \Psi (\tr(A\Psi)\geq 0))\rightarrow (\tr
A\rho\geq 0)],
\end{eqnarray}
where $A\in\hermops$ and $\Psi$ is a pure product state. To see
this, note that the subformula enclosed in square brackets means
``$-A$ is not an entanglement witness for $\rho$'', so that if
this statement is true for all $A$ then there exists no
entanglement witness detecting $\rho$. When $[\rho]$ is rational,
our experience in Section \ref{subsec_EisertsEtalApproach} with
polynomial constraints tells us that the formula in
(\ref{eqn_FOLogicalStatementOfSep}) can be written in terms of
``quantified polynomial inequalities'' with rational coefficients:
\begin{eqnarray}\label{eqn_FOLS}
\forall X   \lbrace (\forall Y \left[ Q(Y)\rightarrow (r(X,Y)\geq
0)\right])\rightarrow (s(X)\geq 0) \rbrace,
\end{eqnarray}
where
\begin{itemize}
\item $X$ is a block of real variables parametrizing the matrix
$A\in\hermops$ (with respect to an orthogonal rational Hermitian
basis of $\hermops$); the ``Hermiticity'' of $X$ is hard-wired by
the parametrization;

\item $Y$ is a block of real variables parametrizing the matrix
$\Psi$;

\item $Q(Y)$ is a conjunction of four polynomial equations that
are equivalent to the four constraints $\tr((\tr_{j}(\Psi))^2)=1$
and $\tr((\tr_{j}(\Psi))^3)=1$ for $j\in\{\A,\B\}$;

\item $r(X,Y)$ is a polynomial representing the expression
$\tr(A\Psi)$;\footnote{To ensure the Hermitian basis is rational,
we do not insist that each of its elements has unit Euclidean
norm. If the basis is $\{X_i\}_{i=0,1,\ldots,\dima^2\dimb^2}$,
where $X_0$ is proportional to the identity operator, then we can
ignore the $X_0$ components write
$A=\sum_{i=1}^{\dima^2\dimb^2}A_iX_i$ and
$\Psi=\sum_{i=1}^{\dima^2\dimb^2}\Psi_iX_i$.  An expression for
$\tr(A\Psi)$ in terms of the real variables $A_i$ and $\Psi_i$ may
then look like $\sum_{i=1}^{\dima^2\dimb^2}A_i\Psi_i\tr(X_i^2)$.}

\item $s(X)$ is a polynomial representing the expression
$\tr(A[\rho])$.
\end{itemize}
The main point of Tarski's result is that the quantifiers (and
variables) in the above sentence can be eliminated so that what is
left is just a formula of elementary algebra involving Boolean
connections of atomic formula of the form $(\alpha \diamond 0)$
involving terms $\alpha$ consisting of rational numbers, where
$\diamond$ stands for any of $<, >, =, \neq$; the truth of the
remaining (very long) formula can be computed in a straightforward
manner.  The best algorithms for deciding (\ref{eqn_FOLS}) require
a number of arithmetic operations roughly equal to
$(PD)^{O(|X|)\times O(|Y|)}$, where $P$ is the number of
polynomials in the input, $D$ is the maximum degree of the
polynomials, and $|X|$ ($|Y|$) denotes the number of variables in
block $X$ ($Y$) \cite{BPR96}. Since $P=6$ and $D=3$, the running
time is roughly $2^{O(\dima^4\dimb^4)}$.

\subsubsection{Approximate formulations}\label{subsubsec_ApproxFormulations}

The benefit of EXACT QSEP is that, compared to Definition
\ref{def_FormalQuSep}, it eliminated any uncertainty in the input
by disallowing irrational matrix entries. Consider the following
motivation for an alternative to EXACT QSEP, where, roughly, we
only ask whether the input $[\rho]$ corresponds to something
\emph{close to} separable:
\begin{itemize}
\item Suppose we really want to determine the separability of a
density operator $\rho$ such that $[\rho]$ has irrational entries.
If we use the EXACT QSEP formulation (so far, we have no decidable
alternative), we must first find a rational approximation to
$[\rho]$.  Suppose the (Euclidean) distance from $[\rho]$ to the
approximation is $\delta$.  The answer that the Tarski-style
algorithm gives us might be wrong, if $\rho$ is not more than
$\delta$ away from the boundary of $\sep$.

\item Suppose the input matrix came from measurements of many
copies of a physical state $\rho$.  Then we only know $[\rho]$ to
some degree of approximation.

\item The best known Tarski-style algorithms for EXACT QSEP
have gigantic running times. 
 Surely, we can achieve better asymptotic running
times if we use an approximate formulation.
\end{itemize}
\noindent Thus, in many cases of interest, insisting that an
algorithm says exactly whether the input matrix corresponds to a
separable state is a waste of time. In Section
\ref{subsubsec_DefnInNP}, we will see that there is another reason
to use an approximate formulation, if we would like the problem to
fit nicely in the theory of NP-completeness.

Gurvits was the first to use the weak membership formulation of
the quantum separability problem \cite{GLS88, Gur03}.  For
$x\in\mathbf{R}^n$ and $\delta>0$, let $B(x,\delta):=
\{y\in\mathbf{R}^n: ||x-y||\leq\delta\}$.  For a convex subset
$K\subset\mathbf{R}^n$, let $S(K,\delta):=\cup_{x\in K}
B(x,\delta)$ and $S(K,-\delta):=\{x: B(x,\delta)\subseteq K\}$.

\vspace*{12pt}
\begin{definition}[Weak membership problem for $K$ (WMEM($K$))]\label{def_WMEM}
Given a rational vector $p\in\mathbf{R}^n$ and rational
$\delta>0$, assert either that
\begin{eqnarray}
p&\in& S(K,\delta), \hspace{2mm}\textrm{or}\label{eqn_WMEMSepAssertion}\\
p&\notin& S(K,-\delta)\label{eqn_WMEMEntAssertion}.
\end{eqnarray}
\end{definition}

\vspace*{12pt} \noindent Denote by WMEM($\sep$) the quantum
separability problem formulated as the weak membership problem. An
algorithm solving WMEM($\sep$) is a separability test with
two-sided ``error'' in the sense that it may assert
(\ref{eqn_WMEMSepAssertion}) when $p$ represents an entangled
state and may assert (\ref{eqn_WMEMEntAssertion}) when $p$
represents a separable state.  Any formulation of the quantum
separability problem will have (at least) two possible answers --
one corresponding to ``$p$ approximately represents a separable
state'' and the other corresponding to ``$p$ approximately
represents an entangled state''. Like in WMEM($\sep$), there may
be a region of $p$ where both answers are valid. We can use a
different formulation where this region is shifted to be either
completely outside $\sep$ or completely inside $\sep$:
\vspace*{12pt}
\begin{definition}[In-biased
weak membership problem for $K$
(WMEM$_\textrm{In}$($K$))]\label{def_WMEMS} Given a rational
vector $p\in\mathbf{R}^n$ and rational $\delta>0$, assert either
that
\begin{eqnarray}
p&\in& S(K,\delta), \hspace{2mm}\textrm{or}\label{eqn_WMEMSSepAssertion}\\
p&\notin& K\label{eqn_WMEMSEntAssertion}.
\end{eqnarray}
\end{definition}
\vspace*{12pt}
\begin{definition}[Out-biased weak membership problem for $K$ (WMEM$_\textrm{Out}$($K$))]\label{def_WMEME}
Given a rational vector $p\in\mathbf{R}^n$ and rational
$\delta>0$, assert either that
\begin{eqnarray}
p&\in& K, \hspace{2mm}\textrm{or}\label{eqn_WMEMESepAssertion}\\
p&\notin& S(K,-\delta)\label{eqn_WMEMEEntAssertion}.
\end{eqnarray}
\end{definition}
\vspace*{12pt} \noindent We can also formulate a ``zero-error''
version such that when $p$ is in such a region, then any algorithm
for the problem has the option of saying so, but otherwise must
answer exactly: \vspace*{12pt}
\begin{definition}[Zero-error weak membership problem for $K$ (WMEM$^0$($K$))]\label{def_WMEM0}
Given a rational vector $p\in\mathbf{R}^n$ and rational
$\delta>0$, assert either that
\begin{eqnarray}
p&\in& K, \hspace{2mm}\textrm{or}\label{eqn_WMEM0SepAssertion}\\
p&\notin& K, \hspace{2mm}\textrm{or}\label{eqn_WMEM0EntAssertion}\\
p&\in& S(K,\delta)\setminus
S(K,-\delta)\label{eqn_WMEM0BoundaryAssertion}
\end{eqnarray}
\end{definition}
\vspace*{12pt}

All the above formulations of the quantum separability problem are
based on the Euclidean norm and use the isomorphism between
$\hermops$ and $\mathbf{R}^{\n}$. We could also make similar
formulations based on other operator norms in $\hermops$.  In the
next section, we will see yet another formulation of an entirely
different flavour.  While each formulation is slightly different,
they all have the property that in the limit as the error
parameter approaches 0, the problem coincides with EXACT QSEP\@.
Thus, despite the apparent inequivalence of these formulations, we
recognize that they all basically do the same job. In fact,
WMEM$(\sep)$, WMEM$_\textrm{In}(\sep)$, WMEM$_\textrm{Out}(\sep)$,
and WMEM$(\sep)^0$ are equivalent: given an algorithm for one of
the problems, one can solve an instance $(\rho,\delta)$ of any of
the other three problems by just calling the given algorithm at
most twice (with various parameters).\footnote{To show this
equivalence, it suffices to show that given an algorithm for
WMEM$(\sep)$, one can solve WMEM$_\textrm{Out}(\sep)$ with one
call to the given algorithm (the converse is trivial); a similar
proof shows that one can solve WMEM$_\textrm{In}(\sep)$ with one
call to the algorithm for WMEM$(\sep)$. The other relationships
follow immediately.  Let $(\rho,\delta)$ be the given instance of
WMEM$_\textrm{Out}(\sep)$. Define
$\rho_0:=\rho+\delta(\rho-I_{\dima,\dimb})/2$ and
$\delta_0:=\delta/(2\sqrt{\dima\dimb(\dima\dimb-1)})$. Call the
algorithm for WMEM$(\sep)$ with input $(\rho_0,\delta_0)$. Suppose
the algorithm asserts $\rho_0\notin S(\sep,-\delta_0)$. Then,
because $||\rho-\rho_0||=\frac{\delta}{2}||\rho-I_{\dima,\dimb}||$
and $||\rho-I_{\dima,\dimb}||\leq 1$, we have $\rho\notin
S(\sep,-(\delta_0+\delta/2))$ hence $\rho\notin S(\sep,-\delta)$.
Otherwise, suppose the algorithm asserts $\rho_0\in
S(\sep,\delta_0)$. By way of contradiction, assume that $\rho$ is
entangled.  But then, by convexity of $\sep$ and the fact that
$\sep$ contains the ball
$B(I_{\dima,\dimb},{1}/{\sqrt{\dima\dimb(\dima\dimb-1)}})$, we can
derive that the ball $B(\rho_0,\delta_0)$ does not intersect
$\sep$.  But this implies $\rho_0\notin S(\sep,\delta_0)$ -- a
contradiction.  Thus, $\rho\in\sep$. This proof is a slight
modification of the argument given in \cite{Lut05}.  See also
Lemma 4.3.3 in \cite{GLS88}.}

\subsection{Computational complexity}\label{subsec_CompComplexity}

This section addresses how the quantum separability problem fits
into the framework of complexity theory. I assume the reader is
familiar with concepts such as \emph{problem}, \emph{instance} (of
a problem), \emph{(reasonable, binary) encodings},
\emph{polynomially relatedness}, \emph{size} (of an instance),
\emph{(deterministic and nondeterministic) Turing machine}, and
\emph{polynomial-time algorithm}; all of which can be found in any
of \cite{GJ79, Pap94, NC00}.

Generally, the weak membership problem is defined for a class
$\mathcal{K}$ of convex sets.  For example, in the case of
WMEM($\sep$), this class is $\{\sep\}_{M,N}$ for all integers
$\dima$ and $\dimb$ such that $2\leq\dima\leq\dimb$.  An instance
of WMEM thus includes the specification of a member $K$ of
$\mathcal{K}$.  The size of an instance must take into account the
size $\langle K \rangle$ of the encoding of $K$.  It is reasonable
that $\langle K \rangle\geq n$ when $K\in\mathbf{R}^n$, because an
algorithm for the problem should be able to work efficiently (in
time that is upper-bounded by a polynomial in the size of an
instance) with points in $\mathbf{R}^n$. But the complexity of $K$
matters, too. For example, if $K$ extends (doubly-exponentially)
far from the origin (but contains the origin) then $K$ may contain
points that require large amounts of precision to represent;
again, an algorithm for the problem should be able to work with
such points efficiently (for example, it should be able to add
such a point and a point close to the origin, and store the result
efficiently).  In the case of WMEM($\sep$), the size of the
encoding of $\sep$ may be taken as $\dimb$ (assuming
$\dima\leq\dimb$), as $\sep$ is not unreasonably long or
unreasonably thin: it is contained in the unit sphere in
$\mathbf{R}^{\n-1}$ and contains a ball of separable states of
radius $\Omega(1/\poly(\dimb))$ (see Section
\ref{sec_OneSidedTestsAndRestrictions}).\footnote{Recall that a
function $f(n)$ is in $\Omega(g(n))$ when there exist constants
$c$ and $n_0$ such that $cg(n)\leq f(n)$ for all $n>n_0$.} Thus,
the total size of an instance of WMEM($\sep$), or any formulation
of the quantum separability problem, may also be taken to be
$\dimb$ plus the size of the encoding of $(\rho, \delta)$.

\subsubsection{Review of NP-completeness}\label{subsubsec_ReviewNPCness}

Complexity theory, and, particularly, the theory of
NP-completeness, pertains to \emph{decision problems} -- problems
that pose a yes/no question.  Let $\Pi$ be a decision problem.
Denote by $\D_\Pi$ the set of instances of $\Pi$, and denote the
yes-instances of $\Pi$ by $\Y_\Pi$. Recall that the complexity
class P (respectively, NP) is the set of all problems the can be
decided by a deterministic Turing machine (respectively,
nondeterministic Turing machine) in polynomial time. The following
equivalent definition of NP is perhaps more intuitive:
\vspace*{12pt}
\begin{definition}[NP] A decision problem $\Pi$ is in NP if there exists
a deterministic Turing machine $T_\Pi$ such that for every
instance $I\in \Y_\Pi$ there exists a string $C_I$ of length
$|C_I|\in O(\poly(|I|))$ such that $T_\Pi$, with inputs $C_I$ and
(an encoding of) $I$, can check that $I$ is in $\Y_\Pi$ in time
$O(\poly(|I|))$.
\end{definition}
\vspace*{12pt} \noindent  The string $C_I$ is called a
\emph{(succinct) certificate}.  Let $\Pi^c$ be the complementary
problem of $\Pi$, i.e. $\D_{\Pi^c}\equiv \D_{\Pi}$ and
$\Y_{\Pi^c}:=\D_\Pi \setminus \Y_\Pi$. The class co-NP is thus
defined as $\{\Pi^c: \hspace{2mm}\Pi\in\NP\}$.

Let us briefly review the different notions of ``polynomial-time
reduction'' from one problem $\Pi'$ to another $\Pi$. Let
$\mathcal{O}_\Pi$ be an oracle, or black-boxed subroutine, for
solving $\Pi$, to which we assign unit complexity cost. A
\emph{(polynomial-time) Turing reduction} from $\Pi'$ to $\Pi$ is
any polynomial-time algorithm for $\Pi'$ that makes calls to
$\mathcal{O}_\Pi$.  Write $\Pi'\leq_\T\Pi$ if $\Pi'$ is
Turing-reducible to $\Pi$. A \emph{polynomial-time
transformation}, or \emph{Karp reduction}, from $\Pi'$ to $\Pi$ is
a Turing reduction from $\Pi'$ to $\Pi$ in which $\mathcal{O}_\Pi$
is called at most once and at the end of the reduction algorithm,
so that the answer given by $\mathcal{O}_\Pi$ is the answer to the
given instance of $\Pi'$. In other words, a Karp reduction from
$\Pi'$ to $\Pi$ is a polynomial-time algorithm that (under a
reasonable encoding) takes as input an (encoding of an) instance
$I'$ of $\Pi'$ and outputs an (encoding of an) instance $I$ of
$\Pi$ such that $I'\in \Y_{\Pi'}\Leftrightarrow I\in \Y_{\Pi}$.
Write $\Pi'\leq_\K\Pi$ if $\Pi'$ is Karp-reducible to $\Pi$.  Karp
and Turing reductions are on the extreme ends of a spectrum of
polynomial-time reductions; see \cite{LLS75} for a comparison of
several of them.

Reductions between problems are a way of determining how hard one
problem is relative to another.  The notion of NP-completeness is
meant to define the hardest problems in NP\@.  We can define
NP-completeness with respect to any polynomial-time reduction; we
define \emph{Karp-NP-completeness} and
\emph{Turing-NP-completeness}:
\begin{eqnarray}
\NPCK &:=& \{\Pi\in\NP:\hspace{2mm} \Pi'\leq_\K\Pi \textrm{ for
all
$\Pi'\in\NP$ }\}\\
\NPCT &:=& \{\Pi\in\NP:\hspace{2mm} \Pi'\leq_\T\Pi \textrm{ for
all $\Pi'\in\NP$ }\}.
\end{eqnarray}
We have $\NPCK\subseteq\NPCT$.  Let $\Pi$, $\Pi'$, and $\Pi''$ be
problems in NP, and, furthermore, suppose $\Pi'$ is in $\NPCK$. If
$\Pi'\leq_\T \Pi$, then, in a sense, $\Pi$ is at least as hard as
$\Pi'$ (which gives an interpretation of the symbol
``$\leq_\T$''). Suppose $\Pi'\leq_\T \Pi$ but suppose also that
$\Pi'$ is not Karp-reducible to $\Pi$. If $\Pi'\leq_\K \Pi''$,
then we can say that ``$\Pi''$ is at least as hard as $\Pi$'',
because, to solve $\Pi'$ (and thus any other problem in NP),
$\mathcal{O}_\Pi$ has to be used at least as many times as
$\mathcal{O}_{\Pi''}$; if any Turing reduction proving
$\Pi'\leq_\T \Pi$ requires more than one call to
$\mathcal{O}_\Pi$, then we can say ``$\Pi''$ is harder than
$\Pi$''. Therefore, if $\NPCK\neq\NPCT$, then the problems in
$\NPCK$ are harder than the problems in $\NPCT\setminus\NPCK$;
thus $\NPCK$ are the hardest problems in NP (with respect to
polynomial-time reductions).

A problem $\Pi$ is \emph{NP-hard} when $\Pi'\leq_T\Pi$ for some
Karp-NP-complete problem $\Pi'\in\NPCK$.  The term ``NP-hard'' is
also used for problems other than decision problems. For example,
let $\Pi'\in\NPCK$; then WMEM($\sep$) is NP-hard if there exists a
polynomial-time algorithm for $\Pi'$ that calls
$\mathcal{O}_{\mathrm{WMEM}(\sep)}$.

\subsubsection{Quantum separability problem in
NP}\label{subsubsec_DefnInNP}

Fact \ref{fact_FiniteDecompOfSepState} suggests that the quantum
separability problem is in NP: a nondeterministic Turing machine
guesses $\{(p_i, [\ket{\psi^\A_i}],
[\ket{\psi^\B_i}])\}_{i=1}^{\n}$,\footnote{I use square brackets
to denote a matrix with respect to the standard basis.}~ and then
easily checks that
\begin{eqnarray}\label{eqn_NPCheck}
[\rho]=\sum_{i=1}^{\n}p_i
[\ket{\psi^\A_i}][\bra{\psi^\A_i}]\otimes
[\ket{\psi^\B_i}][\bra{\psi^\B_i}].
\end{eqnarray}

\begin{figure} [htbp]
\vspace*{13pt}
\centerline{\epsfig{file=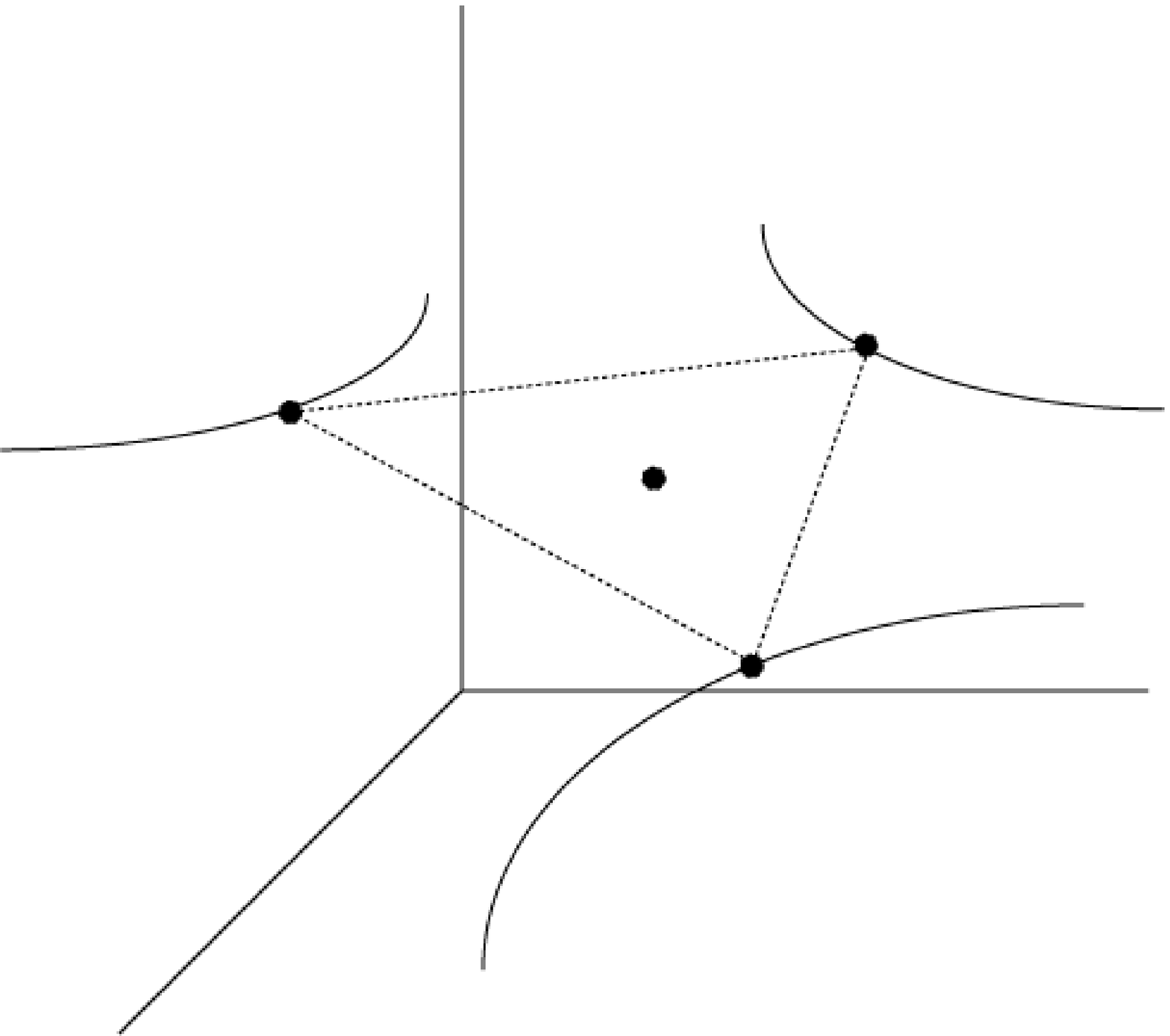, width=8.2cm}} 
\vspace*{13pt} \fcaption{\label{QSEPNotInNP}The dashed triangle
outlines the convex hull of $x_1$, $x_2$, and $x_3$, shown as dots
at the triangle's vertices.  This convex hull contains $\rho$,
shown as a dot inside the triangle, and forms a (schematic) face
of $\sep$. The curves represent the allowable choices for the
$\tilde{x}_i$. Sliding any of the $x_i$ takes
$\conv\{x_1,x_2,x_3\}$ outside of the face.  Incidentally, $\sep$
has no maximum-dimensional faces (facets); this follows from
results in \cite{GL06}.}
\end{figure}

Technically, membership in NP is only defined for decision
problems. Since none of the weak membership formulations of the
quantum separability problem can be rephrased as decision problems
(because problem instances corresponding to states near the
boundary of $\sep$ can satisfy both possible answers), we cannot
consider their membership in NP (but see Section
\ref{subsubsec_NonmembershipInCoNP}, where we define NP for
promise problems). However, EXACT QSEP \emph{is} a decision
problem. \vspace*{12pt}
\begin{problem}
Is \emph{EXACT QSEP} in \emph{NP}?
\end{problem}
\vspace*{12pt} \noindent Hulpke and Bru{\ss} \cite{HB05} have
formalized some important notions related to this problem.  They
show that if $\rho\in S(\sep,-\delta)$, for some $\delta>0$,
then each of the extreme points $x_i\in\sep$ in the expression
$\rho=\sum_{i=1}^{\dima^2\dimb^2}p_ix_i $ can be replaced by
$\tilde{x}_i$, where $[\tilde{x}_i]$ has rational entries.  This
is possible because the extreme points (pure product states) of
$\sep$ with rational entries are dense in the set of all extreme
points of $\sep$. However, when $\rho\notin S(\sep,-\delta)$, then
this argument breaks down. For example, when $\rho$ has full rank
and is on the boundary of $\sep$, then ``sliding'' $x_i$ to a
rational position $\tilde{x}_i$ might cause $\tilde{x}_i$ to be
outside of the affine space generated by $\{x_i\}_{i=1,\ldots,k}$.
Figure \ref{QSEPNotInNP} illustrates this in $\mathbf{R}^3$.
\noindent Furthermore, even if $x_i$ can be nudged comfortably to
a rational $\tilde{x}_i$, one would have to prove that
$\langle\tilde{x}_i\rangle\in O(\poly(\langle[\rho]\rangle))$,
where $\langle X\rangle$ is the size of the encoding of $X$.

So, either the definition of NP does not apply (for weak
membership formulations), or we possibly run into problems near
the boundary of $\sep$ (for exact formulations).  Below we give an
alternative formulation that is in NP; we will refer to this
problem as QSEP\@.  The definition of QSEP is just a precise
formulation of the question ``Given a density operator $\rho$,
does there exist a separable density operator $\hat{\sigma}$ that
is close to $\rho$?''  \vspace*{12pt}
\begin{definition}[QSEP] Given a rational density matrix $[\rho]$ of dimension
$MN$-by-$MN$, and positive rational numbers $\delta_p$,
$\epsilon'$ and $\delta'$; does there exist a distribution
$\{(\tilde{p}_i;
\tilde{\alpha}_i,\tilde{\beta}_i)\}_{i=1,2,...,\n}$ of
unnormalized pure states $\tilde{\alpha}_i\in\mathbf{C}^M$,
$\tilde{\beta}_i\in\mathbf{C}^N$ where $\tilde{p}_i\geq 0$, and
$\tilde{p}_i$ and all elements of $\tilde{\alpha}_i$ and
$\tilde{\beta}_i$ are $\lceil \log_2(1/\delta_p)\rceil$-bit
numbers (complex elements are $x+iy$, $x,y\in\mathbf{R}$; where
$x$ and $y$ are $\lceil \log_2(1/\delta_p)\rceil$-bit numbers)
such that
\begin{eqnarray}\label{QSEP_Requirement1}
\left|1-
||\tilde{\alpha}_i||^2||\tilde{\beta}_i||^2\sum_{j=1}^{\n}
\tilde{p}_j\right| < \epsilon'\hspace{5mm}\textrm{for all $i$}
\end{eqnarray}
and
\begin{eqnarray}\label{QSEP_Requirement2}
||[\rho] -\tilde{\sigma}||^2_2:= \tr(([\rho] -\tilde{\sigma})^2) <
\delta'^2,
\end{eqnarray}
where $\tilde{\sigma}:= \sum_{i=1}^{\n} \tilde{p}_i
\tilde{\alpha}_i\tilde{\alpha}_i^\dagger\otimes
\tilde{\beta}_i\tilde{\beta}_i^\dagger $?
\end{definition}
\vspace*{12pt} \noindent Note that these checks can be done
exactly in polynomial-time, as they only involve elementary
arithmetic operations on rational numbers. To reconcile this
definition with the above question, we define $\hat{\sigma}$ as
the separable density matrix that is the ``normalized version'' of
$\tilde{\sigma}$:
\begin{eqnarray}\label{eqn_DefnOfSigmaHat}
\hat{\sigma} := \sum_{i=1}^{\n} \hat{p}_i
\hat{\alpha}_i\hat{\alpha}_i^\dagger\otimes
\hat{\beta}_i\hat{\beta}_i^\dagger,
\end{eqnarray}
where $\hat{p}_i:=\tilde{p}_i/\sum_i \tilde{p}_i$,
$\hat{\alpha}_i:=\tilde{\alpha}_i/||\tilde{\alpha}_i||$, and
$\hat{\beta}_i:=\tilde{\beta}_i/||\tilde{\beta}_i||$.  Using the
triangle inequality, we can derive that
\begin{eqnarray}
||\hat{\sigma}-\tilde{\sigma}||_2 \leq \sum_{i}\hat{p}_i \left|1-
||\tilde{\alpha}_i||^2||\tilde{\beta}_i||^2\sum_j
\tilde{p}_j\right|,
\end{eqnarray}
where the righthand side is less than $\epsilon'$ when
(\ref{QSEP_Requirement1}) is satisfied.  If
(\ref{QSEP_Requirement2}) is also satisfied, then we have
\begin{eqnarray}
||[\rho]-\hat{\sigma}||_2 \leq ||[\rho] -\tilde{\sigma}||_2+
||\hat{\sigma}-\tilde{\sigma}||_2 \leq \delta' + \epsilon',
\end{eqnarray}
which says that the given $[\rho]$ is no further than $\delta' +
\epsilon'$ away from a separable density matrix (in Euclidean
norm).\footnote{I have formulated these checks to avoid division;
this makes the error analysis of the next section simpler.}

The decision problem QSEP is trivially in NP, as a
nondeterministic Turing machine need only guess the $\lceil\log_2(
1/\delta_p)\rceil$-bit distribution $\{(\tilde{p}_i;
\tilde{\alpha}_i,\tilde{\beta}_i)\}_{i=1,2,...,\n}$ and verify (in
polytime) that (\ref{QSEP_Requirement1}) and
(\ref{QSEP_Requirement2}) are satisfied.

\subsubsection{NP-Hardness}\label{subsubsec_NPHardnessOfQSEP}

Gurvits has shown WMEM($\sep$) to be NP-hard (with respect to the
complexity-measures $M$ and  $\langle\delta\rangle$, i.e.
$\min\{M,N\}$ and $1/\delta$ must be allowed to increase)
\cite{Gur03}. More details about this result appear in Section
\ref{subsubsec_StrongNPHness}.


We check now that QSEP is NP-hard, by way of a Karp-reduction from
WMEM($\sep$).  We assume we are given an instance
$I:=([\rho],\delta)$ of WMEM($\sep$) and we seek an instance
$I':=([\rho'],\delta_p,\epsilon',\delta')$ of QSEP such that if
$I'$ is a ``yes''-instance of QSEP, then $I$ satisfies
(\ref{eqn_WMEMSepAssertion}); otherwise $I$ satisfies
(\ref{eqn_WMEMEntAssertion}). It suffices to use $[\rho']=[\rho]$.
It is clear that if $\delta'$ and $\epsilon'$ are chosen such that
$\delta \geq \delta'+\epsilon'$, then $I'$ is a ``yes''-instance
only if $I$ satisfies (\ref{eqn_WMEMSepAssertion}).  For the other
implication, we need to bound the propagation of some
truncation-errors.  Let $p:=\lceil\log_2( 1/\delta_p)\rceil$.

Recall how absolute errors accumulate when multiplying and adding
numbers.  Let $x=\tilde{x}+\Delta_x$ and $y=\tilde{y}+\Delta_y$
where $x$, $y$, $\tilde{x}$, $\tilde{y}$, $\Delta_x$, and
$\Delta_y$ are all real numbers.  Then we have
\begin{eqnarray}\label{eqn_errorMult}
xy &=& \tilde{x}\tilde{y} + \tilde{x}\Delta_y+ \tilde{y}\Delta_x +
\Delta_x\Delta_y\\
x+y &=& \tilde{x}+\tilde{y}+\Delta_x+\Delta_y.
\end{eqnarray}
For $|\tilde{x}|,|\tilde{y}|<1$, because we will be dealing with
summations of products with errors, it is sometimes convenient
just to use
\begin{eqnarray}\label{eqn_errorMult}
|xy -\tilde{x}\tilde{y}| &\leq&   |\Delta_y|+ |\Delta_x| +
\textrm{max}\{|\Delta_x|,|\Delta_y|\}
\end{eqnarray}
to obtain our cumulative errors (which do not need to be tight to
show NP-hardness).  For example, if $\tilde{x}$ and $\tilde{y}$
are the $p$-bit truncations of $x$ and $y$, where $|x|,|y|<1$,
then $|\Delta_x|, |\Delta_y|<2^{-p}$; thus a conservative bound on
the error of $\tilde{x}\tilde{y}$ is
\begin{eqnarray}
|xy- \tilde{x}\tilde{y}| < |\Delta_y|+|\Delta_x|+|\Delta_x| =
3|\Delta_x|<2^2|\Delta_x|= 2^{-(p-2)}.
\end{eqnarray}

\begin{proposition}\label{Prop_BoundDist__Sigma_SigmaTilde} Let $\sigma\in \sep$ be such that
$\sigma=\sum_{i=1}^{\n}p_i\alpha_i\alpha_i^\dagger\otimes\beta_i\beta_i^\dagger$,
and let \\ $\{(\tilde{p}_i;
\tilde{\alpha}_i,\tilde{\beta}_i)\}_{i=1,2,...,\n}$ be the $p$-bit
truncation of $\{({p}_i;
{\alpha}_i,{\beta}_i)\}_{i=1,2,...,\n}$.\\
Then $||\sigma-\tilde{\sigma}||_2< M^3N^32^{-(p-7.5)}$, where
\begin{eqnarray}
\tilde{\sigma} :=
\sum_{i=1}^{\n}\tilde{p}_i\tilde{\alpha}_i\tilde{\alpha}_i^\dagger\otimes\tilde{\beta}_i\tilde{\beta}_i^\dagger.
\end{eqnarray}
\end{proposition}
\vspace*{12pt} \noindent \textbf{Proof} Letting
$\gamma_i:=p_i\alpha_i\alpha_i^\dagger\otimes\beta_i\beta_i^\dagger
-
\tilde{p}_i\tilde{\alpha}_i\tilde{\alpha}_i^\dagger\otimes\tilde{\beta}_i\tilde{\beta}_i^\dagger$,
we use the triangle inequality to get
\begin{eqnarray}
||\sigma  - \tilde{\sigma}||_2 &\leq& \sum_i ||\gamma_i||_2
=\sum_i \sqrt{\tr(\gamma_i^2)}.
\end{eqnarray}
It suffices to bound the absolute error on the elements of
$[\tilde{p}_i\tilde{\alpha}_i\tilde{\alpha}_i^\dagger\otimes\tilde{\beta}_i\tilde{\beta}_i^\dagger]$;
using our conservative rule (\ref{eqn_errorMult}), these elements
have absolute error less than $2^{-(p-7)}$.  Thus $[\gamma_i]$ is
an $MN$-by-$MN$ matrix with elements no larger than $2^{-(p-7)}$
in absolute value.  It follows that $(\tr(\gamma_i^2))^{1/2}$ is
no larger than $\sqrt{MN}2^{-(p-7.5)}$ in absolute value. Finally,
we get
\begin{eqnarray}
||\sigma - \tilde{\sigma}||_2 \leq\sum_i
\sqrt{\tr(\gamma_i^2)}\leq M^{3}N^{3}2^{-(p-7.5)}.\square
\end{eqnarray}

\vspace*{12pt}
\begin{proposition}\label{Prop_BoundOnNormalisedSigma}
Let $\tilde{\sigma}$ be as in Proposition
\ref{Prop_BoundDist__Sigma_SigmaTilde}.  Then for all
$i=1,2,\ldots \n$
\begin{eqnarray}
\left|1-
||\tilde{\alpha}_i||^2||\tilde{\beta}_i||^2\sum_{j=1}^{\n}
\tilde{p}_j\right| < \dima^3\dimb^3 2^{-(p-5)}.
\end{eqnarray}
\end{proposition}

\vspace*{12pt}\noindent\textbf{Proof} The absolute error on
$\sum_j \tilde{p}_j$ is $\n2^{-p}$. The absolute error on
$||\tilde{\alpha}_i||^2$ (resp. $||\tilde{\beta}_i||^2$) is no
more than $M2^{-(p-3)}$ (resp. $N2^{-(p-3)}$).  This gives total
absolute error of
\begin{eqnarray}
\left|1-{||\tilde{\alpha}_i||^2||\tilde{\beta}_i||^2}\sum_j
\tilde{p}_j\right| < \dima^3\dimb^3 2^{-(p-5)}.\square
\end{eqnarray}
\vspace*{12pt}

Let $\delta':= M^{3}N^{3}2^{-(p-8)}$ and
$\epsilon':=\dima^3\dimb^3 2^{-(p-5)}$ and set $p$ such that
$\epsilon'+\delta'\leq\delta$. Suppose there exists a separable
density matrix $\sigma$ such that $||[\rho]-\sigma||_2=0$.  Then
Propositions \ref{Prop_BoundDist__Sigma_SigmaTilde} and
\ref{Prop_BoundOnNormalisedSigma} say that there exists a
certificate $\tilde{\sigma}$ such that (\ref{QSEP_Requirement1})
and (\ref{QSEP_Requirement2}) are satisfied. Therefore, if $I'$ is
a ``no''-instance, then for all separable density matrices
$\sigma$, $||[\rho]-\sigma||_2>0$; which implies that $I$
satisfies (\ref{eqn_WMEMEntAssertion}).  This concludes a polytime
Karp-reduction from WMEM($\sep$) to QSEP (actually, from
$\textrm{WMEM}_{\textrm{In}}(\sep)$ to QSEP):\vspace*{12pt}
\begin{fact}\label{Prop_QSEPisNPHard}
\emph{QSEP} is in $\NPCT$.
\end{fact}

\subsubsection{Nonmembership in
co-NP}\label{subsubsec_NonmembershipInCoNP}

Technically, WMEM($\sep$) is not in NP because it is not a
decision problem; but it is a promise problem.  Recall that a
\emph{promise problem} $\Pi$ may be defined as a generalization of
a decision problem, where, instead of just yes-instances $\Y_\Pi$
and no-instances $\No_\Pi$, we allow a third set of
maybe-instances (the ``promise'' is that the given instance is in
$\Y_\Pi \cup \No_\Pi$). For $\Pi= \textrm{WMEM($\sep$)}$, we have
$\Y_{\Pi} = \{(x,\delta): x\in S(\sep,-\delta),\delta>0\}$ and
$\No_\Pi = \{(x,\delta): x\in \densops\setminus
S(\sep,\delta),\delta>0\}$ (where we have implicitly restricted
all states to being rational density matrices $[\rho]$). For our
purposes, a promise problem $\Pi$ is defined to be in
\emph{Promise-NP} if every yes-instance has a succinct certificate
of being a yes-instance or a maybe-instance. Accordingly,
WMEM($\sep$) is clearly in Promise-NP.

Is either EXACT QSEP or QSEP in co-NP?   To avoid possible
technicalities, we might first consider the presumably easier
question of whether WMEM($\sep$) is in Promise-co-NP: Does every
entangled state $\rho\notin S(\sep,\delta)$ have a succinct
certificate of not being in $S(\sep,-\delta)$?  It may or may not
be the case that P equals NP$\cap$co-NP, but a problem's
membership in NP$\cap$co-NP can be ``regarded as suggesting'' that
the problem is in P \cite{GJ79}. Thus, we might believe that
WMEM($\sep$) is not in Promise-co-NP (since WMEM($\sep$) is
NP-hard).

Let us consider this with regard to entanglement witnesses, which
are candidates for succinct certificates of entanglement. We know
that every entangled state has an entanglement witness
$A\in\hermops$ that detects it (see footnote on page
\pageref{footnote_DefnEW}). However, it follows from the
NP-hardness of WMEM($\sep$) and Theorem 4.4.4 in \cite{GLS88} that
the weak validity problem for $K=\sep$ (WVAL($\sep$)) is
NP-hard:\footnote{Theorem 4.4.4 in \cite{GLS88}, applied to
$\sep$, states that there exists an oracle-polynomial-time
algorithm that solves the WSEP($\sep$) given an oracle for
WVAL($\sep$).}\vspace*{12pt}
\begin{definition}[Weak validity problem for $K$ (WVAL($K$))]\label{def_WVAL}
Given a rational vector $c\in\mathbf{R}^n$, a rational number
$\gamma$, and rational $\epsilon>0$, assert either that
\begin{eqnarray}
c^Tx &\leq& \gamma+\epsilon\textrm{ for all }x\in K, \hspace{2mm}\textrm{or}\label{WVAL_AssertIsASepPlane}\\
c^Tx &\geq& \gamma-\epsilon \textrm{ for some }x\in K.
\end{eqnarray}
\end{definition}\vspace*{12pt}
\noindent So there is no known way to check efficiently that a
hyperplane $\pi_{A,b}$ separates $\rho$ from $\sep$ (given just
the hyperplane); thus, an entanglement witness alone does not
serve as a succinct certificate of a state's entanglement unless
WVAL($\sep$) is polytime solvable.  However, one could imagine
that there is a succinct certificate of the fact that a hyperplane
$\pi_{A,b}$ separates $\rho$ from $\sep$. If such a certificate
exists, then WVAL($\sep$) is in Promise-NP (and thus WMEM($\sep$)
is in Promise-co-NP).\footnote{WVAL(K) is in Promise-NP if, for
any instance $c$, $\gamma$, $\epsilon$ satisfying
$c^Tx\leq\gamma-\epsilon\textrm{ for all }x\in K$, there exists a
succinct certificate of the fact that $c^Tx \leq
\gamma+\epsilon\textrm{ for all }x\in K$.}

With regard to QSEP, we have the following:\vspace*{12pt}
\begin{fact}\label{Fact_QSEPincoNPmeansNPiscoNP} QSEP is not in co-NP, unless NP equals co-NP.
\end{fact}\vspace*{12pt}
\noindent This follows from the fact that if any
Turing-NP-complete problem is in co-NP, then NP equals co-NP
\cite{Pap94}.
It is strongly conjectured that NP and co-NP are different
\cite{Pap94}, thus we might believe that QSEP is not in co-NP\@.
We would like to be able to use Fact
\ref{Fact_QSEPincoNPmeansNPiscoNP} to show that WVAL($\sep$) is
not in Promise-NP unless NP equals co-NP\@. However, for this, we
would require that WVAL($\sep$) is in Promise-NP implies QSEP is
in co-NP; but this is not the case, because exhibiting a
separating hyperplane for $\rho$ (i.e. showing that $\rho$ is
entangled) does not make $\rho$ a no-instance of QSEP.

\subsubsection{Strong NP-hardness}\label{subsubsec_StrongNPHness}

The NP-complete problem known as PARTITION may be defined as
follows: Given a nonnegative integral vector $a\in \mathbf{Z}^n$,
does there exist a solution $z\in \{-1,1\}^n$ to the equation
$a^Tz=0$? It is well known that there exists a ``dynamic
programming'' algorithm that solves PARTITION in time $O(\poly(n
||a||_1))$, where $||a||_1$ is the sum of the elements of $a$
\cite{GJ79}. This is known as a \emph{pseudopolynomial-time
algorithm}, because \emph{if} $a$ is restricted such that $||a||_1
\in O(\poly (n))$, \emph{then} the algorithm runs in ``polynomial
time''.

Aaronson \cite{Aar05} notes that Gurvits' original NP-hardness
result (in \cite{Gur03}) more precisely shows that WMEM($\sep$) is
NP-hard provided that $1/\delta$ is exponentially large, as I
briefly explain now. For this section only, we switch convention:
$M\geq N$. The full reduction chain that Gurvits uses to prove
NP-hardness is
\begin{equation}
\textrm{PARTITION} \leq_\K \textrm{RSDF} \leq_\K
\textrm{WVAL}(\sep) \leq_\T \textrm{WMEM}(\sep),
\end{equation}
where the robust semidefinite feasibility (RSDF) problem is
defined as follows: \vspace*{12pt}
\begin{definition}[RSDF] Given $k$ $l\times l$, rational, symmetric matrices
$B_1,\ldots, B_k$ and rational numbers $\zeta$ and $\eta$, assert
either that
\begin{eqnarray}
F(B_1,\ldots,B_k)
&\leq& \zeta+\eta, \hspace{2mm}\textrm{or}\\
F(B_1,\ldots,B_k) &\geq& \zeta-\eta,
\end{eqnarray}
where $F(B_1,\ldots,B_k):=\max_{x\in\mathbf{R}^l, ||x||_2=1}
\sum_{i'=1}^k (x^T B_{i'} x)^2$.
\end{definition}
\vspace*{12pt} \noindent  Given a PARTITION instance
$a\in\mathbf{Z}^l$, we want to solve it using an oracle for RSDF.
The reduction (in \cite{BN98}) from PARTITION to RSDF says that
$\eta$ needs to be on the order of $1/\poly(l||a||_2)$.  But this
implies that, for $a$ to be an NP-hard instance of PARTITION,
$1/\eta$ needs to be exponentially large in $l$.  In other words,
WVAL($\sep$) (and hence WMEM($\sep$)) is only shown to be NP-hard
when the accuracy parameter is very small. It is still
conceivable, though, that WVAL($\sep$) (resp. WMEM($\sep$)) is
tractable when $1/\epsilon$ (resp. $1/\delta$) is in
$O(\poly(M,N))$.  Below, I show that a new reduction discovered by
Gurvits \cite{Gur06} (inspired by the proof of Lemma 3 in
\cite{GB05}) removes the possibility for such a family of
WVAL($\sep$) instances in a certain regime of $\epsilon$;
moreover, I also remove this possibility for a problem slightly
more difficult than the weak membership problem for $\sep$.

The new reduction chain is
\begin{equation}
\textrm{CLIQUE} \leq_\K \textrm{WMQS} \leq_\K \textrm{RSDF}
\leq_\K \textrm{WVAL}(\sep) \leq_\T \textrm{WMEM}(\sep),
\end{equation}
where CLIQUE $\in\NPCK$ (see \cite{GJ79}) is the problem of
deciding whether the number of vertices in the largest complete
subgraph (clique) of a given simple graph on $n$ vertices is at
least $c$ (given also integer $c\leq n$), and WMQS is the problem
of weakly deciding a bound on the maximum of the quadratic form
$y^TAy$ over the simplex $\Delta_n :=\{y\in\mathbf{R}^n: y_i\geq
0, ||y||_1=1\}$: \vspace*{12pt}
\begin{definition}[WMQS] Given rational,
symmetric $A\in\mathbf{R}^{n\times n}$ with nonnegative entries
$A_{ij}$ and rational numbers $\zeta'$ and $\eta'>0$, assert
either that
\begin{eqnarray}
H(A)
&\leq& \zeta'+\eta', \hspace{2mm}\textrm{or}\\
H(A) &\geq& \zeta'-\eta',
\end{eqnarray}
where $H(A):=\max_{y\in\Delta_n} y^T A y$.
\end{definition}
\vspace*{12pt}

The first link in this chain is well known via the following
theorem:\footnote{Thanks to Etienne de Klerk at University of
Waterloo for pointing me to this theorem.} \vspace*{12pt}
\begin{mytheorem}[\cite{MS65}]\label{thm_MAXCLIQUESIMPLEX}
Let $G$ be a simple graph on $n$ vertices, and let $A_G$ be the
adjacency matrix for $G$.\footnote{$A_G$ has a 1 in position
$(i,j)$ whenever $(i,j)$ is an edge of $G$, and otherwise has a 0
($A_G$ has just zeros on the diagonal).}~ Let $\kappa$ be the size
of the maximum complete subgraph of $G$. Then
\begin{eqnarray}
\max_{y\in\Delta_n} y^T A_G y = 1-1/\kappa.
\end{eqnarray}
\end{mytheorem}
\vspace*{12pt} \noindent Suppose $(G,c)$ is a given CLIQUE
instance, where $G$ has $n$ vertices.  To transform $(G,c)$ into a
WMQS-instance, just set $\zeta'$ to be the midpoint of interval
$I_c=[1-1/(c-1), 1-1/c]$ and set $\eta'$ so that the interval
$[\zeta'-\eta', \zeta'+\eta']$ is strictly contained in $I_c$ (and
set $A:=A_G$). Note that such an $\eta'$ exists in
$\Omega((c-1)^{-1} - c^{-1}) \in \Omega(n^{-2})$. Therefore, WMQS
is NP-hard (with respect to $n$ and $\langle \eta' \rangle$) even
when $1/\eta'$ is restricted to being in $O(\poly(n))$, which we
call \emph{strong} NP-hardness (see \cite{GJ79}).

The second link Gurvits establishes by noting that, via a change
of variables $y_i \rightarrow x_i^2$,
\begin{eqnarray}
\max_{y\in\Delta_n} y^T A y = \max_{x\in\mathbf{R}^n,||x||_2=1}
\sum_{i,j=1}^n A_{ij}x_i^2x_j^2,
\end{eqnarray}
and
\begin{eqnarray}
\sum_{i,j=1}^n A_{ij}x_i^2x_j^2 = \sum_{i,j=1}^n
(\sqrt{A_{ij}}x_ix_j)^2 = 2\sum_{1\leq i<j\leq
n}(\sqrt{A_{ij}}x_ix_j)^2 =\sum_{1\leq i<j\leq n} (x^TB^{ij}x)^2,
\end{eqnarray}
where $B^{ij}$, for all $1\leq i<j\leq n$, is the matrix with
$\sqrt{A_{ij}}$ in positions $(i,j)$ and $(j,i)$, and zeros
elsewhere (note there are $n(n-1)/2$ matrices $B^{ij}$). Thus RSDF
is strongly NP-hard (with respect to $l$ and $\langle\eta\rangle$)
in the regime $k\geq l(l-1)/2$, because we can make some of the
blocks $B_{i'}$ zero-blocks (the rest of the
instance-transformation is $(\zeta, \eta):=(\zeta',\eta')$).

The third link is established in \cite{Gur03}, where Gurvits shows
that, for
\begin{eqnarray}
B:= \begin{pmatrix} 0 & B_1 & \cdots & B_{M-1}   \\
B_1 & 0 & \cdots &0\\
\vdots & \vdots & \ddots & \vdots \\
B_{M-1} & 0 & \cdots & 0
 \end{pmatrix},
\end{eqnarray}
where the zeros are $N\times N$ blocks of 0 and the $B_i$ are real
symmetric $N\times N$ matrices, the following holds:
\begin{eqnarray}
\max_{\sigma\in\sep} \tr(B\sigma) =
\max_{x\in\mathbf{R}^N,||x||_2=1}\sum_{i'=1}^{M-1}(x^T B_{i'}
x)^2.
\end{eqnarray}
It follows that WVAL($\sep$) is strongly NP-hard (with respect to
$N$ and $\langle\epsilon\rangle$) in the regime $M \geq
N(N-1)/2+1$ (again, the rest of the instance-transformation is
trivial: $(\gamma,\epsilon):=(\zeta,\eta)$). But suppose we had an
oracle for WVAL($\sep$)$_{N \leq M \leq N(N-1)/2}$ and wanted to
solve the instance of CLIQUE. Then, by setting $M:=N:=n(n-1)/2 +1$
and making each $B_{i'}$ block ($i'=1,\ldots,N-1$) zeros but for
the upper left $n\times n$ submatrix (into which we put $B^{ij}$),
we have that
\begin{eqnarray}
\max_{x\in\mathbf{R}^N,||x||_2=1}\sum_{i'=1}^{M-1}(x^T B_{i'} x)^2
=\max_{x\in\mathbf{R}^n,||x||_2=1}\sum_{1\leq i<j\leq n}
(x^TB^{ij}x)^2.
\end{eqnarray}
Thus WVAL($\sep$) is also strongly NP-hard (with respect to $N$
and $\langle\epsilon\rangle$) in the regime $2\leq N\leq M$.

The last link in the reduction is more correctly split up into two
links:
\begin{equation}
\textrm{WVAL}(\sep) \leq_\K \textrm{WVIOL}(\sep) \leq_\T
\textrm{WMEM}(\sep),
\end{equation}
where the following definition applies:

\vspace*{12pt}
\begin{definition}[Weak violation problem for $K$ ($\textrm{WVIOL($K$)}$)]\label{def_WVAL}
Given a rational vector $c\in\mathbf{R}^n$, a rational number
$\gamma$, and rational $\epsilon>0$, either
\begin{itemize}
\item \textrm{assert $c^Tx \leq \gamma+\epsilon$ for all $x\in K$, or}\label{WVAL_AssertIsASepPlane}\\
\item \textrm{find a vector $y\in S(K,\epsilon)$ with $c^Ty \geq
\gamma - \epsilon$}.
\end{itemize}
\end{definition}
\vspace*{12pt}

\noindent It is clear that WVIOL($\sep$) is also strongly NP-hard.
But the Turing-reduction from WVIOL($\sep$) to WMEM($\sep$) is
highly nontrivial in that the only proof of this reduction,
appearing in Theorem 4.3.2 of \cite{GLS88}, requires the
shallow-cut ellipsoid method.  The accuracy parameters for the
WMEM($\sep$)-oracle queries in this reduction only have
exponentially small lower bounds.  Thus the problem remains:

\vspace*{12pt}
\begin{problem} Is \emph{WMEM}$(\sep)$ tractable when $1/\delta$ is in
$O(\poly(M,N))$?
\end{problem}
\vspace*{12pt}

Let us consider the following problem, which is more difficult
than the weak membership problem because it asks for the normal
vector to a separating hyperplane in the case where the given
point is not inside the convex set: \vspace*{12pt}
\begin{definition}[Weak separation problem for $K$ ($\textrm{WSEP($K$)}$)]\label{def_WSEP}
Given a rational vector $p\in\mathbf{R}^n$ and rational
$\delta>0$, either
\begin{itemize}
\item \textrm{assert $p\in S(K,\delta)$,
or}\label{eqn_WSEPSepAssertion}\\
\item \textrm{find a rational vector $c\in\mathbf{R}^n$ with
$||c||_\infty= 1$ such that $c^Tx < c^Tp+\delta$ for every $x\in
K$.}\label{eqn_WSEPEntAssertion}
\end{itemize}
\end{definition}
\vspace*{12pt} \noindent  Note that WSEP($\sep$) asks either to
assert that the given density matrix is almost separable, or to
find an approximate entanglement witness.  A Turing reduction from
WVIOL($\sep$) to WSEP($\sep$) (one of which appears in Theorem
4.2.2 of \cite{GLS88}), is much more straightforward and does not
require the ellipsoid method -- any cutting-plane algorithm for
the weak nonemptiness problem\footnote{The weak nonemptiness
problem for $K$ is merely to find a point in $S(K,\epsilon)$ or
assert that $S(K,-\epsilon)$ is empty.}~ relative to a weak
separation oracle suffices. Applying the analytic-center algorithm
of Atkinson and Vaidya \cite{AV95} gives a Turing-reduction that
only needs to make polynomially-accurate WSEP($\sep$)-queries:

\vspace*{12pt}
\begin{fact}
\emph{WSEP}$(\sep)$ is strongly NP-hard (w.r.t. $N$ and
$\langle\delta\rangle$) in the regime $2\leq N\leq M$.
\end{fact}
\vspace*{12pt}

\subsubsection{Possibility of a Karp Reduction}\label{subsubsec_TowardsKarp}

To date, every decision problem (except for QSEP) that is known to
be in $\NPCT$ is also known to be in $\NPCK$
\cite{PS01}.\footnote{By ``known to be in $\NPCT$'', I mean that
the language corresponding to the decision problem can be defined
and shown to be Turing-NP-complete, independent of any unproven
assumptions. See \cite{Pav03} for languages that are suspected to
be Turing-but-not-Karp-NP-complete, whose existence depends on
unproven assumptions about NP.}~~ While it is strongly suspected
that Karp and Turing reductions are inequivalent within NP, it
would be surprising if QSEP, or some other formulation of the
quantum separability problem,\footnote{By ``formulation of the
quantum separability problem'', I mean an NP-contained approximate
formulation that tends to EXACT QSEP as the accuracy parameters of
the problem tend to zero.}~ is the first example that proves this
inequivalence: \vspace*{12pt}
\begin{problem}\label{prob_QSEPinNPCK}
Is \emph{QSEP} in $\NPCK$?
\end{problem}\vspace*{12pt}
\noindent One reason to believe that a Turing reduction is
necessary to prove the NP-hardness of the quantum separability
problem is that a proof (see Section
\ref{subsubsec_StrongNPHness}) seems to require a reduction from
$\textrm{WVAL}(\sep)$ to $\textrm{WMEM}(\sep)$; in turn, this
reduction seems to require the shallow-cut ellipsoid method (a
Turing reduction). It is a long-standing open problem as to
whether the reduction from $\textrm{WVAL}(\sep)$ to
$\textrm{WMEM}(\sep)$ can be done differently.  However, as the
NP-hardness of WMEM($\sep$) is a relatively recent result, there
may be an altogether different proof of it that does not require a
reduction from $\textrm{WVAL}(\sep)$ to $\textrm{WMEM}(\sep)$.

Note that, because of Fact \ref{Prop_QSEPisNPHard}, a negative
answer to Problem \ref{prob_QSEPinNPCK} implies that $\P\neq\NP$.
Note also that a direct reduction from some $\Pi'\in\NPCK$ to QSEP
(or some other formulation) would depend heavily on the precise
definition of QSEP rather than the true spirit of the quantum
separability problem captured by WMEM($\sep$).  Thus, if the
answer to Problem \ref{prob_QSEPinNPCK} is positive, it might be
easier to look for a Karp-reduction from some $\Pi \in \NPCK$ to
WMEM($\sep$).

\section{Survey and complexity analysis of algorithms for the quantum separability
problem}\label{sec_SurveyOfAlgorithms}

For the survey, I concentrate on proposed algorithms that solve an
approximate formulation of the quantum separability problem and
have (currently known) asymptotic analytic bounds on their running
times.  For this reason, the SDP relaxation algorithm of Eisert et
al.\ is not mentioned here (see Section
\ref{subsec_EisertsEtalApproach}); though, I do not mean to
suggest that in practice it could not outperform the following
algorithms on typical instances.  As well, I do not analyze the
complexity of the naive implementation of every necessary and
sufficient criterion for separability, as it is presumed that this
would yield algorithms of higher complexity than the following
algorithms.  For an exhaustive list of all such criteria, see the
book by Bengtsson and Zyczkowski \cite{BZ06}.

I give complexity estimates for several of the algorithms
surveyed. The main purpose below is to get a time-complexity
estimate in terms of the parameters $\dima$, $\dimb$, and
$\delta$, where $\delta$ is the accuracy parameter in
WMEM($\sep$).
The running-time estimates are based on the number of elementary
arithmetic operations and do not attempt to deal with computer
round-off error; I do not give estimates on the total amount of
machine precision required.  Instead, where rounding is necessary
in order to avoid exponential blow-up of the representation of
numbers during the computation, I assume that the working
precision\footnote{``Working precision'' is defined as the number
of significant digits the computer uses to represent numbers
during the computation.}~ can be set large enough that the overall
effect of the round-off error on the final answer is either much
smaller than $\delta$ or no larger than, say, $\delta/2$ (so that
doubling $\delta$ takes care of the error due to round-off).

\subsection{Naive algorithm and $\delta$-nets}\label{sec_BasicAlgorithm}

The naive algorithm for any problem in NP consists of a search
through all potential succinct certificates that the given problem
instance is a ``yes''-instance.  Thus QSEP immediately gives an
algorithm for the quantum separability problem.   Hulpke and
Bru{\ss} \cite{HB05} have demonstrated another hypothetical
guess-and-check procedure that does not involve the probabilities
$p_i$. They noticed that, given the vectors $\{[\ket{\psi^\A_i}],
[\ket{\psi^\B_i}]\}_{i=1}^{\n}$, one can check that
\begin{eqnarray}\label{eqn_NPCheckHulpkeBruss1}
\textrm{$\{[\ket{\psi^\A_i}][\bra{\psi^\A_i}]\otimes
[\ket{\psi^\B_i}][\bra{\psi^\B_i}]\}_{i=1}^{\n}$ is affinely
independent; and}
\end{eqnarray}
\begin{eqnarray}\label{eqn_NPCheckHulpkeBruss2}
[\rho]\in\conv \{[\ket{\psi^\A_i}][\bra{\psi^\A_i}]\otimes
[\ket{\psi^\B_i}][\bra{\psi^\B_i}]\}_{i=1}^{\n}
\end{eqnarray}
in polynomially many arithmetic operations (and they give an
algorithm for the separability problem based on this observation).
We can, in principle, reformulate QSEP to incorporate the ideas of
Hulpke and Bru{\ss} in order to get a better naive algorithm. The
reader is referred to \cite{qphIoa05} for details (and for how our
approach differs from theirs -- essentially, their algorithm
solves a more exact formulation of the separability problem); we
quote the asymptotic running-time estimate of
$(MN/\delta)^{O(M^3N^2 +M^2N^3)}$.

QSEP can be further reformulated to avoid searching through
\emph{all} $p$-bit-precise pure states by using the concept of a
net on (or covering of) the unit sphere.  Let $\mathbf{S}_\dima$
be the Euclidean unit sphere in $\mathbf{C}^M$. In principle, we
can use a \emph{Euclidean $\delta$-net of $\mathbf{S}_\dima$},
which is a minimal set of points
$\mathcal{N}^M_{\delta}:=\{\ket{x_i}\}_{i=1}^{|\mathcal{N}^M_{\delta}|}\subset
\mathbf{S}_\dima$ such that for any $\ket{x}\in\mathbf{S}_\dima$
there exists $\ket{x_i}\in \mathcal{N}^M_{\delta}$ such that
$||\ket{x}-\ket{x_i}||_2\leq\delta$.  The optimal size of a
$\delta$-net on the real sphere is known to be no larger than $(1
+ 2/\delta)^{M}$ \cite{P89}, thus we can take
$|\mathcal{N}^M_{\delta}|\leq (1 +
2/\delta)^{2M}$.\footnote{Recall that the Euclidean distance
between two vectors in $\mathbf{C}^M$ depends only on the real
part of their inner product, which behaves exactly like the
dot-product of real vectors in $\mathbf{R}^{2M}$.}~ Assuming the
availability of asymptotically optimal $\mathcal{N}^M_{\delta}$
for all $M$ and $\delta$ (where the real elements in each
$\ket{x_i}$ have $p$ bits of precision), the complexity of the
naive algorithm for separability is reduced to
$(2/\delta)^{O(M^3N^2+M^2N^3)}$, which simply corresponds to the
number of $M^2N^2$-subsets of $\mathcal{N}^M_{\delta} \times
\mathcal{N}^N_{\delta}$.  We will assume availability, or
\emph{advice}, of $\delta$-nets for the complexity estimates of
several of the following algorithms.

\subsection{Bounded search for symmetric extensions}\label{sec_DohertyEtalKonigRenner}

In Section \ref{sec_DohertyEtalApproach}, we considered two tests
-- one that searches for symmetric extensions of $\rho$, and a
stronger one that searches for PPT symmetric extensions.  Now I
show that recent results can put an upper bound on the number $k$
of copies of subsystem $\A$ when solving an approximate
formulation of the separability problem. The bound only assumes
symmetric extensions, \emph{not} PPT symmetric extensions, so it
is possible that a better bound may be found for the stronger test
(Problem \ref{Problem_CanBoundOnkBeImproved}).

If a symmetric state
$\varrho\in\densopsgen{(\mathbf{C}^d)^{\otimes n}}$ has a
symmetric extension to $\densopsgen{(\mathbf{C}^d)^{\otimes
(n+m)}}$ for all $m>0$, then it is called \emph{(infinitely)
exchangeable}.  The quantum de Finetti theorem (see \cite{DPS04}
and references therein) says that the infinitely exchangeable
state $\varrho$ is separable. Recalling the terminology of Section
\ref{sec_DohertyEtalApproach}, it is also possible to derive that,
for $\rho\in\densopsgen{\CMotimesCN}$, if there exists a symmetric
extension of $\rho$ to $k$ copies of subsystem $\A$ for all $k>0$,
then $\rho\in\sep$.  This is the result that proves that Doherty
et al.'s hierarchy of tests is complete:  if $\rho$ is entangled,
then the SDP at some level $k_0$ of the hierarchy will not be
feasible (i.e. will not find a symmetric extension of $\rho$ to
$k_0$ copies of subsystem $\A$).

It seems reasonable that, if we are only interested in whether
$\rho$ is $\delta$-close to $\sep$, we should not need to check
for extensions of $\rho$ to $k$ copies of subsystem $\A$ for $k$
larger than some bound $\bar{k}=\bar{k}(M,N,\delta)$.  In fact, we
can use results of Christandl et al. \cite{qphCKMR06} to compute
just such a $\bar{k}$.\footnote{Thanks to Andrew Doherty for
calling my (and Christandl et al.'s!) attention to this;
otherwise, I would be deriving a worse bound on $k$, based on
results in \cite{KR05}.}~  We require the following theorem:
\vspace*{12pt}
\begin{mytheorem}[\cite{qphCKMR06}]
Suppose $\rho\in\densops$ and there exists a Bose-symmetric
extension $\rho'$ of $\rho$ to $k\geq 2$ copies of subsystem $\A$,
i.e. $(I\otimes P)\rho'=\rho'$ for all $k!$ permutations $P$ of
the $k$ copies of subsystem $\A$. Then
\begin{eqnarray}
\tr|\rho - \sigma|\leq \frac{4\dima}{{k}},
\end{eqnarray}
for some $\sigma\in\sep$.
\end{mytheorem}
\vspace*{12pt} \noindent Note that the result uses the \emph{trace
distance}, $\tr|X-Y|$, between two operators $X$ and $Y$.  Let us
assume we are solving the weak membership formulation of the
quantum separability problem with respect to the trace distance,
and with accuracy parameter $\delta$.  Then, setting
$\delta={4\dima}/{{k}}$, we get the following upper bound for $k$:
\vspace*{12pt}
\begin{corollary} To solve \emph{WMEM($\sep$)} (with respect to the trace distance)
with accuracy parameter $\delta$ by searching for symmetric
extensions (as described in Section
\ref{sec_DohertyEtalApproach}), it suffices to look for symmetric
extensions to
\begin{eqnarray}
\bar{k}:=\lceil 4\dima/\delta\rceil
\end{eqnarray}
copies of subsystem A.
\end{corollary}
\vspace*{12pt} To estimate the total complexity of the algorithm,
note that
\begin{eqnarray}
d_{S_k} &=& {{M+k-1}\choose{k}} \approx {{M+k}\choose{k}} =
\frac{(M+k)!}{k!M!} \\
&\approx&
\frac{1}{\sqrt{2\pi}}\frac{(M+k)^{M+k}}{k^kM^M}  \left({\frac{1}{k}+\frac{1}{M}}\right)^{1/2},\\
\end{eqnarray}
where in the last line we used Stirling's approximation $n!\approx
n^n\sqrt{2\pi n}/e^n$.  Substituting $\bar{k}\approx 4M/\delta$
for $k$, we get
\begin{eqnarray}\label{eqn_MainDohertEtalKonigRennerComplexity}
\left({\frac{1}{\bar{k}}+\frac{1}{M}}\right)^{-1/2}d_{S_{\bar{k}}}&\approx&
\frac{1}{\sqrt{2\pi}}
\frac{(M+4M/\delta)^{M+\bar{k}}}{(4M/\delta)^{\bar{k}}M^M}
=\frac{1}{\sqrt{2\pi}}
\frac{(1+4/\delta)^{M+\bar{k}}}{(4/\delta)^{\bar{k}}} \\
&\approx& \frac{1}{\sqrt{2\pi}}(4/\delta)^M\\
d_{S_{\bar{k}}}&\approx& \frac{1}{\sqrt{2\pi}}(4/\delta)^M
\left({{\delta}/{{4M}}+{1}/{M}}\right)^{1/2}\\
&\approx& \frac{1}{\sqrt{2\pi}}(4/\delta)^M
\left({1}/{M}\right)^{1/2}.
\end{eqnarray}
Just to solve the first constraint in (\ref{prob_DohertyEtalSDP})
requires $\sqrt{n}$ (but usually far fewer) iterations of a
procedure that requires $O(m^2n^2)$ arithmetic operations, for
$m=((d_{S_{\bar{k}}})^2-\dima^2)\dimb^2$ and
$n=(d_{S_{\bar{k}}})^2\dimb^2$ \cite{DPS04}.  That is, the
complexity of each iteration is on the order of\\
$(d_{S_{\bar{k}}})^8\poly(M,N,\log(1/\delta))$.

\vspace*{12pt}
\begin{problem}\label{Problem_CanBoundOnkBeImproved}
Can the upper bound $\bar{k}$ be improved by taking into
consideration the PPT constraints in (\ref{prob_DohertyEtalSDP})?
\end{problem}
\vspace*{12pt}

\noindent We mention that a larger bound $\bar{k}'\gg \bar{k}$ on
$k$ can be derived from a theorem in \cite{KR05}, which also can
be used to compute an approximate separable decomposition of
$\rho$ in the case where the SDP algorithm finds a symmetric
extension of $\rho$ to $\bar{k}'$ copies of subsystem $\A$.

\subsection{Entanglement-witness search via global optimization}\label{sec_NewSepAlg}

Recall that an \emph{entanglement witness (for $\rho$)} is defined
to be any operator $A\in\hermops$ such that
$\tr(A\sigma)<\tr(A\rho)$ for all $\sigma\in\sep$ and some
$\rho\in\ent$; we say that ``$A$ detects $\rho$''. Since every
$\rho\in\ent$ has an entanglement witness that detects it
\cite{HHH96}, one way to solve the quantum separability problem is
to exhaustively (but not naively!) search for an entanglement
witness for the given $\rho$.  We mention that the dual of the SDP
in the symmetric-extension search algorithm can be used to find an
(approximate) entanglement witness for $\rho$ (when the SDP is
infeasible) \cite{DPS04}.

\subsubsection{Large SDP method}\label{subsubsec_LargeSDPPerezGarcia}

P{\'{e}}rez-Garcia and Cirac \cite{PC06} note that the following
SDP effectively searches for an approximate entanglement witness
$-A$ for $\rho$:
\begin{eqnarray}
&\textrm{minimize} \hspace{2mm}& \tr(A\rho)\\
&\textrm{subject to:}\hspace{2mm}& \bra{x_i}A\ket{x_i}\geq 0,
\textrm{ for all $\ket{x_i}\in\mathcal{N}^M_\delta$, and
$\tr(A)=1$}.
\end{eqnarray}
\noindent Define the following convex hull:
\begin{eqnarray}
C:= \conv \{\ketbra{x_i}{x_i}\otimes\ketbra{b}{b}:
\ket{x_i}\in\mathcal{N}^M_\delta, \ket{b}\in\mathbf{C}^N\}.
\end{eqnarray}
\noindent If the minimum is negative, then $-A$ is an approximate
entanglement witness for $\rho$ because there is a hyperplane with
normal $-A$ that separates $\rho$ from $C$ (otherwise $\rho$ is in
$C$ and is thus in $\sep$).  This is justified because for any
$\rho\in\sep$ there exists a $\sigma\in C$ such that $||\rho -
\sigma||_1 \leq 2\delta$, as we now verify.  First, note that if
$||\ket{x}-\ket{y}||_2 \leq \delta <1$, then
$\textrm{Re}\braket{x}{y} \geq 1-\delta^2/2$ (since
$||\ket{x}-\ket{y}||^2_2=2-2\textrm{Re}\braket{x}{y}$) and thus
\begin{eqnarray}
||\ketbra{x}{x}-\ketbra{y}{y}||_1 &=& 2\sqrt{1-|\braket{x}{y}|^2}\hspace{10mm}\textrm{(see \cite{NC00}, p. 415)} \\
&\leq& 2 \sqrt{1-(\textrm{Re}\braket{x}{y})^2} \\
&\leq& 2 \sqrt{1-(1-\delta^2/2)^2}\\
&=& 2\delta \sqrt{1-\delta^2/4}\\
&<& 2\delta.
\end{eqnarray}
\noindent If $\rho =
\sum_j\lambda_j\ketbra{u_j}{u_j}\otimes\ketbra{v_j}{v_j}$, for
$\lambda_j\geq 0$ and $\sum_j\lambda_j=1$, then choosing
$\ket{x_{i_j}}\in \mathcal{N}^M_\delta$ such that $||\ket{x_{i_j}}
- \ket{u_j}||_2\leq\delta$ makes $\sigma:=
\sum_j\lambda_j\ketbra{x_{i_j}}{x_{i_j}}\otimes\ketbra{v_j}{v_j}$
in $C$ with $||\rho - \sigma||_1 \leq 2\delta$.

The size (up to a constant factor) of the constraint of the SDP is
$n=|\mathcal{N}^M_{\delta}| N$ (and the number of real variables
to parametrize $A$ is $m = M^2N^2-1$), thus the complexity of one
iteration of the SDP is of the order
$|\mathcal{N}^M_{\delta}|^2\poly(M,N,\log(1/\delta))$ (assuming
$\mathcal{N}^M_{\delta}$ is available).

This approach is a discretization of Brand{\~{a}}o and Vianna's
robust semidefinite program (see Section
\ref{subsubsec_OtherTests}), which is
\begin{eqnarray}
&\textrm{minimize} \hspace{2mm}& \tr(A\rho)\\
&\textrm{subject to:}\hspace{2mm}& x^\dagger A x\geq 0, \textrm{
for all $x\in\mathbf{C}^M$, and $\tr(A)=1$}.
\end{eqnarray}
\noindent (Note that, combined with Gurvits' NP-hardness result
\cite{Gur03}, this demonstrates that robust semidefinite programs
are, in general, NP-hard.)  Essentially, P{\'{e}}rez-Garcia and
Cirac have removed the robustness by using a $\delta$-net, which
clarifies the complexity of the approach for deterministically
solving WMEM($\sep$).

\subsubsection{Interior-point cutting-plane algorithm with global optimization
subroutine}\label{subsubsec_IoannouAlgorithm}

The algorithms in \cite{ITCE04,IT06} solve WMEM($\sep$) by solving
$\textrm{WSEP}(\sep)$ using a subroutine for WOPT($\sep$):

\vspace*{12pt}
\begin{definition}[Weak optimization problem for $K$ (WOPT(K))]\label{def_WOPT}
Given a rational vector $c\in\mathbf{R}^n$ and rational
$\epsilon>0$, either
\begin{itemize}
\item find a rational vector $y\in\mathbf{R}^n$ such that $y\in
S(K,\epsilon)$ and  $c^Tx\leq c^Ty +\epsilon$ for every $x\in K$;
or

\item assert that $S(K,-\epsilon)$ is empty.\footnote{For
$\epsilon$ we consider, this will never be the case for us, as we
only consider WOPT($\sep$) and $\sep$ is far from
empty.}\label{eqn_WSEPEntAssertion}
\end{itemize}
\end{definition}
\vspace*{12pt}

\noindent Clearly, using a subroutine for WOPT($\sep$) allows one
to test whether a Hermitian operator $A$ approximately detects
$\rho$. The algorithms effectively perform a binary search through
the space of all entanglement witnesses.  For example, the
algorithm in \cite{ITCE04} searches through the space
$\mathcal{W}:=\{A \in\hermops: \tr(A)=0, \tr(A)\leq 1\}$ of all
normalized entanglement witnesses by iterating the following:
\begin{romanlist}

\item Let $A$ be an approximate center (interior point) of the
current search space (which is initialized to $\mathcal{W}\cap
\{x\in\hermops: \tr(\rho x)\geq 0\}$, but subsequently gets
approximately halved in each iteration, in step (iv) below).

\item Set $A:=A/||A||_2$ and give $(A,\epsilon:=\delta/5)$ to
WOPT($\sep$) subroutine, which outputs $\sigma_A$.

\item If $A$ approximately detects $\rho$, then return $A$;

\item otherwise, use $\sigma_A$ to generate a cutting plane which
approximately halves the current search space (cuts it through $A$
and the origin).  If current search space is too small to contain
a hyperplane that separates $\rho$ from $S(\sep,\delta)$, then
return ``$\rho\in S(\sep,\delta)$''.

\end{romanlist}



The number of arithmetic operations required by the algorithm
described in \cite{ITCE04} is
\begin{eqnarray}
O((T+
\dima^6\dimb^6\log(1/\delta))\dima^2\dimb^2\log^2(\dima^2\dimb^2/\delta)),
\end{eqnarray}
where $T$ is the cost of one call to the WOPT($\sep$) subroutine
(see \cite{qphIoa05} for details).

Now consider the complexity of computing an instance
$(A,\epsilon)$ of WOPT($\sep$), $||A||_2=1$. For each $x_i\in
\mathcal{N}^M_{\delta}$, we compute $\ket{j_i}:=
\textrm{argmax}_{\ket{j}}|\bra{x_i}\bra{j}A\ket{x_i}\ket{j}|$ via
eigenvector analysis of $\bra{x_i}A\ket{x_i}$ (which is
Hermitian).\footnote{Thanks to David P{\'{e}}rez-Garcia for
suggesting this method.}  Let $\tilde{i}$ denote the index $i$ of
an element $\ket{x_i}$ of $\mathcal{N}^M_{\delta}$ that maximizes
$|\bra{x_i}\bra{j_i}A\ket{x_i}\ket{j_i}|$.  Then
$\ket{x_{\tilde{i}}}\ket{j_{\tilde{i}}}$ may be taken as a
solution to WOPT($\sep$) because
\begin{eqnarray}\label{ineq_netmaximizer}
|\bra{x_{\tilde{i}}} \bra{j_{\tilde{i}}} A  \ket{x_{\tilde{i}}}
\ket{j_{\tilde{i}}}
-\max_{\ket{\alpha}\ket{\beta}}\bra{\alpha}\bra{\beta}A\ket{\alpha}\ket{\beta}|
\leq 2\delta,
\end{eqnarray}
as we now verify.  Let $\ket{a}\ket{b}:=
\textrm{argmax}_{\ket{\alpha}\ket{\beta}}\bra{\alpha}\bra{\beta}A\ket{\alpha}\ket{\beta}$
and let $i^*$ denote the index $i$ of an element $\ket{x_i}$ of
$\mathcal{N}^M_{\delta}$ such that $||\ket{x_i}-\ket{a}||_2 \leq
\delta$.  Writing $\ket{yb}$ for $\ket{y}\ket{b}$ for any
$\ket{y}$,
\begin{eqnarray}
&&|\bra{x_{i^*}b}A\ket{x_{i^*}b} - \bra{ab}A\ket{ab}| \\
&=& |\bra{x_{i^*}b}A\ket{x_{i^*}b} - \bra{ab}A\ket{x_{i^*}b}+
\bra{ab}A\ket{x_{i^*}b} - \bra{ab}A\ket{ab}| \\
&=& ||\ket{x_{i^*}}-\ket{a}||_2\left|
\frac{\bra{x_{i^*}}-\bra{a}}{||\ket{x_{i^*}}-\ket{a}||_2}\bra{b}A\ket{x_{i^*}b}
+
\bra{ab}A\frac{\ket{x_{i^*}}-\ket{a}}{||\ket{x_{i^*}}-\ket{a}||_2}\ket{b}\right|\\
&\leq&
2\delta\max_{\ket{c},\ket{d} \in \mathbf{C}^M}|\bra{c}\bra{b}A\ket{d}\ket{b}|\\
&\leq& 2\delta\max_{\ket{c'},\ket{d'} \in \mathbf{C}^M\otimes\mathbf{C}^N}|\bra{c'}A\ket{d'}|\\
&=& 2\delta\max\{\sqrt{\lambda}: \textrm{$\lambda$ an eigenvalue
of $A^\dagger A$}\}\hspace{10mm}\textrm{(see \cite{HJ85}, p. 312)}\\
&\leq& 2\delta \sqrt{\sum_i \lambda_i(A^\dagger A)}\hspace{10mm}\textrm{(where $\lambda_i(X)$ denotes eigenvalues of $X$)} \\
&=& 2\delta ||A||_2 \hspace{10mm}\textrm{(since $A$ is normal, see \cite{HJ85}, p. 316)}\\
&=& 2\delta.
\end{eqnarray}
\noindent The inequality (\ref{ineq_netmaximizer}) follows from
noting
\begin{eqnarray}
\bra{x_{i^*}}\bra{b}A\ket{x_{i^*}}\ket{b} \leq
\bra{x_{i^*}}\bra{j_{i^*}}A\ket{x_{i^*}}\ket{j_{i^*}} \leq
\bra{x_{\tilde{i}}}\bra{j_{\tilde{i}}}A\ket{x_{\tilde{i}}}\ket{j_{\tilde{i}}}
\leq \bra{a}\bra{b}A\ket{a}\ket{b}.
\end{eqnarray}

Therefore, the complexity of the whole algorithm is on the order
of \\$|\mathcal{N}^M_{\delta}| \poly(M,N,\log(1/\delta))$
(assuming $\mathcal{N}^M_{\delta}$ is available).

\subsection{Other algorithms}

We mention two other algorithms, whose running times cannot be
easily compared to that of the above algorithms.

\subsubsection{Cross-norm criterion via linear
programming}\label{sec_Rudolf&PerezGarcia}

Rudolph \cite{Rud00} derived a simple characterization of
separable states in terms of a computationally complex operator
norm $||\cdot||_\gamma$.  For a finite-dimensional vector space
$V$, let $\mathcal{T}(V)$ be the class of all linear operators on
$V$. The norm is defined on $\mathcal{T}(\mathbf{C}^\dima) \otimes
\mathcal{T}(\mathbf{C}^\dimb)$ as
\begin{eqnarray}
||t||_\gamma := \inf \lbrace \sum_{i=1}^{k}||u_i||_1 ||v_i
||_1:\hspace{1mm} t = \sum_{i=1}^{k}u_i\otimes v_i \rbrace,
\end{eqnarray}
where the infimum is taken over all decompositions of $t$ into
finite summations of elementary tensors, and
$||X||_1:=\tr(\sqrt{X^\dagger X})$.  Rudolph showed that
$||\rho||_\gamma \leq 1$ if and only if $||\rho||_\gamma = 1$, and
that a state $\rho$ is separable if and only if $||\rho||_\gamma =
1$.

P{\'{e}}rez-Garcia \cite{Per04} showed that approximately
computing this norm can be reduced to a linear program (which is a
special case of a semidefinite program): $\min\{ c^Tx :
\hspace{2mm}Ax=b, x\geq 0\}$,
where $A\in\mathbf{R}^{n\times m}$, $b\in\mathbf{R}^n$,
$c\in\mathbf{R}^m$, and $x$ is a vector of $m$ real variables;
here, $x\geq 0$ means that all entries in the vector are
nonnegative.  An LP can be solved in $O(m^{3}L')$ arithmetic
operations, where $L'$ is the length of the binary encoding of the
LP \cite{Ye97}.
The linear program has on the order of $\dima^2\dimb^2$ variables
and $|\mathcal{N}^M_{1/k}|^2|\mathcal{N}^N_{1/k}|^2$ constraints,
where $k$ is an integer that determines the relative
error\footnote{The \emph{relative error} of an approximation
$\tilde{x}$ of $x$ is defined as $|x-\tilde{x}|/x$.}~
$(k/(k-1))^4-1$ on the computation of the norm.  Thus the
complexity of the whole algorithm is on the order of
$|\mathcal{N}^M_{1/k}|^2|\mathcal{N}^N_{1/k}|^2
\poly(M,N,\log(1/\delta))$ (assuming availability of Euclidean
$(1/k)$-nets).

Suppose $|| \rho ||_\gamma$ is found to be no greater than
$1+\eta$. Then, we would like to use $\eta$ to upper-bound the
distance, with respect to either trace or Euclidean norm, from
$\rho$ to $\sep$.  Unfortunately, we do not know how to do this.
This drawback, along with the fact that the error on the computed
norm is relative as opposed to absolute, does not allow this
algorithm to be easily compared to the other algorithms we
consider. 

\subsubsection{Fixed-point iterative method}\label{sec_Zapatrin}

Zapatrin \cite{qphZap05c} suggests an iterative method that solves
the separability problem.\footnote{Facts about iterative methods:
First, the basic Newton-Raphson method in one variable. Suppose
$\xi$ is a zero of a function $f:\mathbf{R}\rightarrow\mathbf{R}$
and that $f$ is twice differentiable in a neighbourhood $U(\xi)$
of $\xi$. Then the Taylor expansion of $f$ about $x_0\in U(\xi)$
gives
\begin{eqnarray}
 0 =f(\xi) &=& f(x_0)+(\xi-x_0)f'(x_0)+ \cdots\\
&=& f(x_0)+(\tilde{\xi}-x_0)f'(x_0),
\end{eqnarray}
where $\tilde{\xi}= x_0 -f(x_0)/f'(x_0)$ is an approximation of
$\xi$. Repeating the process, with a truncated Taylor expansion of
$f$ about $\tilde{\xi}$, gives a different approximation
$\tilde{\tilde{\xi}}=\tilde{\xi}-f(\tilde{\xi})/f'(\tilde{\xi})$.
This suggests the iterative method $x_{i+1}=\Phi(x_i)$, for
$\Phi(x):=x-f(x)/f'(x)$.  If $f'(\xi)\neq 0$, the sequence
$(x_i)_i$ converges to $\xi$ if $x_0$ is sufficiently close to
$\xi$. More generally, if $\Phi(x):\mathbf{R}^n\rightarrow
\mathbf{R}^n$ is a contractive mapping on $B(x_0,r)$, then the
sequence $(x_0, \Phi(x_0),\Phi(\Phi(x_0)),\ldots)$ converges to
the unique fixed point in $B(x_0,r)$ (as long as $\Phi(x_0)\in
B(x_0,r)$) \cite{SB02}.}~~  He defines the function
$\Phi:\hermops\rightarrow\hermops$:
\begin{eqnarray}
\Phi(X):= X + \lambda\left( \rho - \int\int
e^{\bra{\psi^\A}\otimes\bra{\psi^\B}X\ket{\psi^\A}\otimes\ket{\psi^\B}}\ketbra{\psi^\A}{\psi^\A}\otimes\ketbra{\psi^\B}{\psi^\B}d\mathbf{S}_\dima
d\mathbf{S}_\dimb\right),
\end{eqnarray}
where $\mathbf{S}_\dima$ and $\mathbf{S}_\dimb$ are the complex
origin-centred unit spheres (containing, respectively,
$\ket{\psi^\A}$ and $\ket{\psi^\B}$), and $\lambda$ is a constant
dependent on the derivative (with respect to $X$) of the quantity
in parentheses ($\lambda$ is chosen so that $\Phi$ is a
contraction mapping). In earlier work \cite{qphZap04, qphZap05a,
qphZap05b}, Zapatrin proves that any state $\sigma$ in the
interior $\sep^\circ$ of $\sep$ may be expressed
\begin{eqnarray}
 \sigma=\int\int
e^{\bra{\psi^\A}\otimes\bra{\psi^\B}X_\sigma\ket{\psi^\A}\otimes\ket{\psi^\B}}\ketbra{\psi^\A}{\psi^\A}\otimes\ketbra{\psi^\B}{\psi^\B}d\mathbf{S}_\dima
d\mathbf{S}_\dimb \in \sep,
\end{eqnarray}
for some Hermitian $X_\sigma$. Thus the function $\Phi$ has a
fixed point $X_\rho=\Phi(X_\rho)$ if and only if
$\rho\in\sep^\circ$. When $\rho\in\sep^\circ$, then a
neighbourhood (containing $0$) in the domain of $\Phi$ can be
found where iterating $X_{i+1}:=\Phi(X_i)$, starting at $X_0:=0$,
will produce a sequence $(X_i)_i$ that converges to $X_\rho$ when
$\rho\in\sep^\circ$, but diverges otherwise.

Each evaluation of $\Phi(X)$ requires $\dima^2\dimb^2/2
+\dima\dimb$ integrations of the form
\begin{eqnarray}
\int\int
e^{\bra{\psi^\A}\otimes\bra{\psi^\B}X\ket{\psi^\A}\otimes\ket{\psi^\B}}
\braket{\textbf{e}^\A_j}{\psi^\A}\braket{\textbf{e}^\B_{j'}}{\psi^\B}
\braket{\psi^\A}{\textbf{e}^\A_k}\braket{\psi^\B}{\textbf{e}^\B_{k'}}
d\mathbf{S}_\dima d\mathbf{S}_\dimb,
\end{eqnarray}
where $\{\textbf{e}^\A_j\}_j$ and $\{\textbf{e}^\B_k\}_k$ are the
standard bases for $\mathbf{C}^\dima$ and $\mathbf{C}^\dimb$.
However, the off-diagonal ($j\neq k$, $j'\neq k'$) integrals have
a complex integrand so are each really two real integrals; thus
the total number of real integrations is $\dima^2\dimb^2$.  Let
$\Xi_\delta$ represent the number of pure states at which the
integrand needs to be evaluated in order to perform each real
numerical integration, in order to solve the overall separability
problem with accuracy parameter $\delta$. Zapatrin shows that the
approximate number of iterations required is upper-bounded by
$2\dimb(\dimb+1)L(\log(1/\delta),\log(\dimb))$, where $L$ is a
bilinear function of its arguments.  The complexity of the entire
algorithm is roughly $\Xi_\delta
\poly(\dima,\dimb,\log(1/\delta))$ (ignoring $\log(\dimb)$
factors). We can use nets on $\mathcal{S}_M$ and $\mathcal{S}_N$
to estimate the complexity of $\Xi_\delta$.  It is clear that the
numerical integration is more complex than solving WOPT($\sep$);
at the very least, it needs to sample points in $\mathcal{S}_N$ as
well as $\mathcal{S}_M$. Thus, I make the reasonable presumption
that $\Xi_\delta \geq
|\mathcal{N}^M_\delta||\mathcal{N}^N_\delta|$.

\subsection{Complexity comparison of
algorithms and practical
considerations}\label{sec_ComplexityComparison}

The complexity estimates in the survey show that the two best
algorithms are the bounded search for symmetric extensions
(Section \ref{sec_DohertyEtalKonigRenner}) and the cutting-plane
entanglement-witness search algorithm (Section
\ref{subsubsec_IoannouAlgorithm}).  The dominant (exponential)
factors in the asymptotic complexity estimates are
\begin{eqnarray}
(d_{S_{\bar{k}}})^8 &\approx& 2^M\times
(1/\delta)^{8M}\hspace{10mm}{\textrm{(symmetric-extensions
search)}} \\|\mathcal{N}^M_{\delta}|&\approx& 2^M\times
(1/\delta)^{2M} \hspace{10mm}{\textrm{(entanglement-witness
search)}}.
\end{eqnarray}
\noindent As a caveat, recall that the estimate for the
entanglement-witness search algorithm assumes the advice of
asymptotically optimal $\mathcal{N}^M_{\delta}$, which in
principle can be precomputed for the $M$ and $\delta$ of interest
\cite{HSS}.


The bounded symmetric-extension search algorithm only experiences
``exponential slow-down'' when $k$ approaches $\bar{k}$. The SDP
relaxation algorithm of Section \ref{subsec_EisertsEtalApproach}
behaves similarly, in that successive SDP relaxations get larger,
while the first SDP relaxation is feasible. Thus, a good strategy
for a deterministic WMEM($\sep$) solver might be to start with
these algorithms (after exhausting all the other efficient
one-sided tests), and proceed until the SDPs get infeasibly large;
then, switch to the entanglement-witness search algorithm, whose
complexity bottleneck is the WOPT($\sep$) subroutine, which has
constant worst-case complexity throughout the execution of the
algorithm, but whose task -- essentially, searching the domain of
the function
$f(\ket{\alpha},\ket{\beta}):=\bra{\alpha}\bra{\beta}A\ket{\alpha}\ket{\beta}$
-- can be parallelized.  Interestingly, the subroutine need not
find a certified optimum of $f$ until its final execution, and
even then only in the case where the algorithm will output
$\rho\in S(\sep,\delta)$ \cite{ITCE04}; thus, comparing the
outputs of executions of a local optimum finder seeded at a few
random points in the domain may suffice for at least some of the
WOPT($\sep$) calls.  Finally, instead of using $\delta$-nets, the
subroutine should benefit from more-sophisticated, continuous
global optimization methods (which may utilize calculus to
eliminate large chunks of the domain of $f$) such as the
semidefinite programming relaxation method of Lasserre
\cite{Las01}, Lipschitz optimization \cite{HP95}, and Hansen's
global optimization algorithm using interval analysis
\cite{Han92}.

\nonumsection{Acknowledgements} \noindent Thanks to Matthias
Christandl, Roger Colbeck, Anuj Dawar, Andrew Doherty, Daniel
Gottesman, Alastair Kay, Michele Mosca, Renato Renner, and Tom
Stace for helpful discussions. Special thanks to David
P{\'{e}}rez-Garcia for his help with Section \ref{sec_NewSepAlg}.
This work was supported by the CESG (UK), NSERC (Canada), ORS
(UK), RESQ (EU grant IST-2001-37559), and EPSRC.



%
%
%

%

\end{document}